\documentclass[aps, prd, twocolumn, showpacs, superscriptaddress, nofootinbib,floatfix]{revtex4-2}
\usepackage{amsmath,mathrsfs,amsbsy,color,graphicx,bm,amsthm,amsfonts}
\usepackage{bbm}
\usepackage{times}
\usepackage{units}
\usepackage{dcolumn}
\usepackage{graphicx}
\usepackage{epsfig}
\usepackage{epstopdf}
\usepackage[colorlinks,linkcolor=red,anchorcolor=green,citecolor=blue,CJKbookmarks=True]{hyperref}
\DeclareMathSymbol{\shortminus}{\mathbin}{AMSa}{"39}
\usepackage{mathrsfs}
\usepackage{braket}
\usepackage{amssymb}
\usepackage{txfonts}
\usepackage{enumitem}
\usepackage{multirow}
\graphicspath{{fihures/},{pics/}} \usepackage{subfigure} 
\usepackage{microtype}
\usepackage[cmyk]{xcolor} 
\usepackage{placeins}

\begin{document}
\title{Simulation of the massless Dirac field in 1+1D curved spacetime}

    \author{Zhilong Liu}
 
    \affiliation{Department of Physics, Key Laboratory of Low Dimensional Quantum Structures and Quantum Control of Ministry of Education, and Synergetic Innovation Center for Quantum Effects and Applications, Hunan Normal
    University, Changsha, Hunan 410081, P. R. China}

    \author{Run-Qiu Yang}
        \email{aqiu@tju.edu.cn ({Corresponding authors})}\affiliation{Center for Joint Quantum Studies and Department of Physics, School of Science,
        Tianjin University, Yaguan Road 135, Jinnan District, 300350 Tianjin, P. R. China}
    
    \author{Heng Fan}
\affiliation{Institute of Physics, Chinese Academy of Sciences, Beijing 100190, China}
        
  \author{Jieci Wang}
        \email{jcwang@hunnu.edu.cn (Corresponding authors)}\affiliation{Department of Physics, Key Laboratory of Low Dimensional Quantum Structures and Quantum Control of Ministry of Education, and Synergetic Innovation Center for Quantum Effects and Applications, Hunan Normal
    University, Changsha, Hunan 410081, P. R. China}

    \begin{abstract}
    Simulating the nature of quantum fields in diverse spacetime backgrounds offers valuable insights for the fundamental comprehension of quantum mechanics and general relativity.  Here we introduce a novel method for mapping the massless Dirac equation in 1+1D curved spacetime to a controllable quantum simulation model, applicable to various observers' perspectives. We perform numerical simulations of Simpson spacetime and calculate tunneling rates in Painlevé and Schwarzschild coordinates, which align closely with theoretical predictions of Hawking radiation. Additionally, we show the transition of Simpson spacetime from a regular black hole to a wormhole as the parameter $``a > r_s"$. This method facilitates the study of spacetime from various coordinate perspectives (observers), providing deeper insights and understanding.
    \end{abstract}

    \pacs{~}

\maketitle
\section{Introduction}
    Quantum simulation has indeed attracted growing attention in recent years, providing a nuanced approach to explore intricate theories like quantum mechanics and general relativity. The quest to unify them has generated a plethora of fascinating phenomena and predictions, such as the Casimir effect \cite{casimir1948attraction}, particle creation \cite{parker1968particle} and the enigmatic Hawking radiation \cite{hawking1974black}. Notably, Hawking radiation challenges the conventional view of black holes as purely absorptive entities, proposing instead that they emit particles in a manner similar to a blackbody. However, the extremely low temperature of this radiation makes it practically undetectable with current observational capabilities. Similarly, the Unruh effect \cite{unruh1976notes}, which predicts that an accelerating observer in a vacuum would perceive a bath of particles, remains elusive due to the immense accelerations required.
    Following the seminal scheme proposed by Unruh in 1981 \cite{unruh1981experimental}, which suggested the use of a “dumb hole” (a fluid flow analog) to simulate black hole dynamics, the challenge of astronomically observing black holes has been transformed into a feasible experimental setup. 
    
    Although this classical experimental setup is unavoidably prone to noise \cite{rousseaux2008observation,weinfurtner2011measurement}, numerous quantum simulation platforms have emerged to address this issue. These platforms include supersonic fluid \cite{braunstein2023analogue,makinen2023rotating,vsvanvcara2024rotating}, Bose-Einstein condensates \cite{garay2000sonic,steinhauer2016observation,eckel2018rapidly,tian2021probing2,tian2022probing}, superconducting circuits \cite{blencowe2020analogue,terrones2021towards}, optical waveguide arrays \cite{liu2014trapping,koke2016dirac}, and ultracold atoms in optical lattices \cite{pedernales2018dirac}. These simulations primarily link the dynamics equations of the experimental setups with their counterparts in various curved spacetimes, necessitating highly precise control over the interactions within these systems.
    The superconducting circuits, with their ability to couple a large number of qubits and provide wide-range tunable couplers \cite{yan2018tunable,sung2021realization}, as well as their mature manufacturing processes, have significantly advanced the field of quantum simulation. This platform has been employed to simulate various exotic effects in curved spacetime, such as black hole radiation \cite{tian2019analogue,shi2023quantum} and wormhole traversal \cite{Carlos2016Mapping}. A conformal metrics method to simulate a random curved spacetime has been proposed on this platform \cite{koke2016dirac,Sabn2012EncodingRP}, and Yang has proposed a mapping in Eddington-Finkelstein coordinates from 1+1 dimensional spherical symmetry metrics onto the XY and Hubbard models \cite{yang2020simulating}, which has been generalized to the three-dimensional case \cite{deger2023ads}. This platform demonstrates significant potential in the fields of quantum simulation and quantum computing. In curved spacetime, the nature of spacetime and quantum effects can lead to different phenomena being observed by different observers. For instance, the infinite redshift effect near the Schwarzschild horizon and the ``thermal bath" effect experienced by accelerated observers. Therefore, simulating quantum behavior in curved spacetime from the perspectives of different observers is of paramount importance.
    \newline \indent In this paper, we introduce a novel method for simulating the dynamics of massless Dirac particles in curved spacetime, applicable from various coordinate perspectives (observers). By introducing general variable transformations, we can map the massless Dirac equation onto simulatable models, specifically the XY model or the Hubbard model. These models can be experimentally investigated using analogous platform present above. This observer-independent method enables us to study quantum effects in curved spacetime from various perspectives, thereby gaining deeper insights.
    \newline \indent The organization of this paper is as follows. In Section \ref{Sec_II}, we introduce the method that maps a massless Dirac equation to the simulable models. In Section \ref{Sec_IIIA}, we numerically investigate the tunneling rate and the process of particles crossing the event horizon. In Section \ref{Sec_IIIB}, we numerically simulate the regular black hole and the process of wormhole traversal. In Section \ref{Sec_IV}, we provide brief concluding remarks. Additional details that are omitted from the main text are provided in Appendix \ref{Appendix}.

\section{Mapping massless Dirac equations to the XY model}\label{Sec_II}
    Simulating quantum field in curved spacetime using controllable quantum systems can significantly deepen our understanding of both general relativity and quantum theory. This approach not only provides robust experimental validation for theoretical predictions but also offers a viable research method for domains that are challenging to directly experiment with or observe. In this section, we present a detailed method for mapping the massless Dirac equation in 1+1D curved spacetime onto a controllable quantum simulation system. This approach supports the investigation of quantum fields in curved spacetime from different perspectives (observers).

    To investigate the Dirac equation within a general coordinate framework, we begin by considering the following form of the static spacetime metric (choosing the signature (-,+) and for simplicity, we set $\hbar=c=k_B=G=1$ throughout this paper, where $\hbar$ is the reduced Planck constant, $c$ is the speed of light, $k_B$ is the Boltzmann constant, and $G$ is the gravitational constant).
    \begin{equation}\label{eq_1}
        \mathrm{d}s^2=-\mathrm{e}^{A(r)}\mathrm{d}t^2+\mathrm{e}^{B(r)}\mathrm{d}r^2,
    \end{equation}
    where $A(r)$ and $B(r)$ are functions that depend exclusively on the spatial coordinate $r$. In 1+1 dimensional curved spacetime, the Dirac equation can be expressed as \cite{1994Dirac,mann1991semiclassical,morsink1991black}
    \begin{equation}
        i\gamma^a e^{\mu}_{\;\; a}\partial_\mu\psi+\frac{i}{2}\gamma^a\frac{1}{\sqrt{-g}}\partial_\mu(\sqrt{-g}e^{\mu}_{\;\; a})\psi-m\psi=0.  \label{eq_2}
    \end{equation}
    Here, $e^{\mu}_{\;\; a}$ represents the vielbein, $g$ denotes the determinant of the metric tensor, and $m$ is the mass of the Dirac particle (which is zero in the massless case). The $\gamma^{a}$ are the Dirac matrices in locally flat spacetime, satisfying the following relation $\left\{\gamma^a,\gamma^b\right\}=2\eta^{ab}$. Throughout this work, we will adopt the representation $\gamma^a = (\sigma_z, i\sigma_y)$, where $\sigma$ are the Pauli matrices. In the massless case, the Dirac spinor $\psi$ can be decomposed into the form $(\phi, -\phi)^{\mathrm{T}}$.
    By selecting the vielbein $e^{\mu}_{\;\; \nu}$ like
    \begin{equation}
        e^{\mu}_{\;\; a}=
        \begin{pmatrix}
            e^{-\frac{A(r)}{2}} & \sqrt{2}e^{-\frac{A(r)}{2}}  \\
            -\sqrt{2}e^{-\frac{B(r)}{2}} & -e^{-\frac{B(r)}{2}} 
        \end{pmatrix},
    \end{equation}
    then the Dirac equation mentioned above can subsequently be expressed in the following form
    \begin{eqnarray}
        \partial_t\phi=-c(r)\partial_r\phi+V(r)\phi , \label{eq_4}
    \end{eqnarray}
    where $c(r)=e^{\frac{A(r)-B(r)}{2}}$ and $V(r)=-A^{\prime}(r)e^{\frac{A(r)-B(r)}{2}}/4$, with $A^{\prime}(r)$ representing derivative with respect to $r$. 
    For this form of differential equation, we introduce the variable substitution: $\phi=Q(r)\omega$, where the transformation factor is determined as
    \begin{equation} \label{eq_tran}
        Q(r)=K\sqrt{c(r)}e^{\int{\frac{V(r)}{c(r)}dr}},
    \end{equation}
    where $K$ is an arbitrary constant, as arbitrary constant factors do not affect the equation (we will set it to 1 in subsequent discussions).
    We note that due to the complex nature of the wave function, the square root factor does not impose any constraints on the transformation. Under this variable substitution, the above equation (5) can be written as
    \begin{equation}
        \partial_t\omega = -\frac{1}{2} \Big\{ c(r)\partial_r\omega + \partial_r\left[ c(r)\omega \right] \Big\}, 
    \end{equation}
    Subsequently, following the methodology outlined in \cite{yang2020simulating}, we discretize the space as $c_n=c(nd)$,$\omega_n(t)=\omega_n(t,nd)$ and apply the corresponding phase factor $\omega=(-1)^ne^{-i\mu t}\tilde{\omega}$, where the chemical potential $\mu$ is a constant. By imposing the relevant quantization conditions and Heisenberg equations of motion, we derive the effective Hamiltonian
    \begin{equation}
        \mathcal{H}=\sum_{n}\left[{-\kappa_n(\hat{a}^\dagger_n\hat{a}_{n-1}+h.c)-\mu\hat{a}^\dagger_n\hat{a}_n}\right],
    \end{equation}
    where the coupling coefficient $\kappa_n\approx\frac{c_{[(n-1/2)d]}}{2d}$. 
    This Hamiltonian can be mapped onto the isotropic XY model by Jordan-Wigner transformation \cite{1970Statistical,barouch1971statistical}. The resulting model can be implemented on simulation platforms, such as the aforementioned superconducting quantum circuits.

    It is noteworthy that this method offers the advantage of enabling the simulation of quantum fields in curved spacetime across different coordinate systems, thereby greatly enhancing our understanding of both curved spacetime and quantum theory. Specifically, we observe that this approach aligns with the method presented in \cite{yang2020simulating} when $V(r) = 0$ and reduces to the form described in \cite{Carlos2016Mapping,Sabn2012EncodingRP} when $c(r)$ is constant.
    
    To demonstrate the applicability of this transformation method to different observers, the following sections will use the Simpson spacetime as an example. We apply this method to simulate Dirac particles from the perspectives of both Schwarzschild and Painlevé observers. Additionally, we numerically simulate the transition of Simpson spacetime from a regular black hole to a wormhole using this approach.
\section{dynamics of particles in Simpson spacetime}\label{Sec_III}
    In this section, we apply the previously described method to simulate the massless Dirac field in Simpson spacetime. Our primary numerical calculations involve simulating the infinite redshift effect in Schwarzschild coordinates and the process of particles crossing the event horizon into the black hole interior in Painlevé coordinates. We also calculate the tunneling spectra of outgoing waves in both coordinate systems and compared them with Hawking radiation \cite{damour1976black}. Furthermore, we illustrate the behavior of particles as the spacetime transitions from a regular black hole to a wormhole and simulate the process of particles traversing the wormhole. To ensure alignment with experimental implementation and adherence to physical significance, Dirichlet boundary conditions will be employed in all subsequent simulations.

    The Simpson spacetime metric can be written as \cite{simpson2019black}
        \begin{equation}
            \mathrm{d}s^2=-f(r)\mathrm{d}t^2+\frac{1}{f(r)}\mathrm{d}r^2,
        \end{equation}
    where $f(r)=1-\frac{r_s}{\sqrt{r^2+a^2}}$,with $r_s$ being the Schwarzschild radius, $r \in(-\infty,+\infty)$ and $a$ $\in[0,+\infty)$. 
    We can calculate the Kretschmann scalar as follows
        \begin{align}
            R_{abcd}R^{abcd} &= \frac{4}{(r^2+a^2)^2} \left\{ \sqrt{r^2+a^2} \left[ 4a^2r_s(r^2-a^2) \right] \right. \nonumber \\
            &\quad + 3a^4(r^2+a^2) \left. + \frac{3}{4}r_s^2(3a^4-4a^2r^2+4r^4) \right\}.
        \end{align}
        It is evident that the Kretschmann scalar diverges only when $a=0$ and $r=0$, which corresponds to the singularity of the Schwarzschild black hole. For $a \neq 0$, the singularity is disappeared, resulting in a black bounce that bounces matter into a future incarnation of the universe. When $a$ exceeds the Schwarzschild radius $( a > r_s)$, the spacetime structure transitions into a traversable wormhole. Furthermore, the sign of the coordinate $r$ distinguishes between two separate universes.
\subsection{Schwarzschild black hole for \texorpdfstring{$a=0$}{}}\label{Sec_IIIA}
\subsubsection{Black hole in Schwarzschild coordinates \texorpdfstring{$\left\{t,r\right\}$}{}}
    Firstly, we consider the case where $a=0$. In this scenario, the metric simplifies to that of a Schwarzschild black hole.
        Due to the two causally disconnected universes are separated at the singularity with similar nature, we only need to study one side of the universe. Then the metric can be written as
        \begin{equation}\label{eq_sch}
            \mathrm{d}s^2=-(1-\frac{r_s}{r})\mathrm{d}t^2+\frac{1}{(1-\frac{r_s}{r})}\mathrm{d}r^2.
        \end{equation}
        \begin{figure}[t]
            \centering{\subfigure{\includegraphics[width=0.49\linewidth,height=0.36\linewidth]{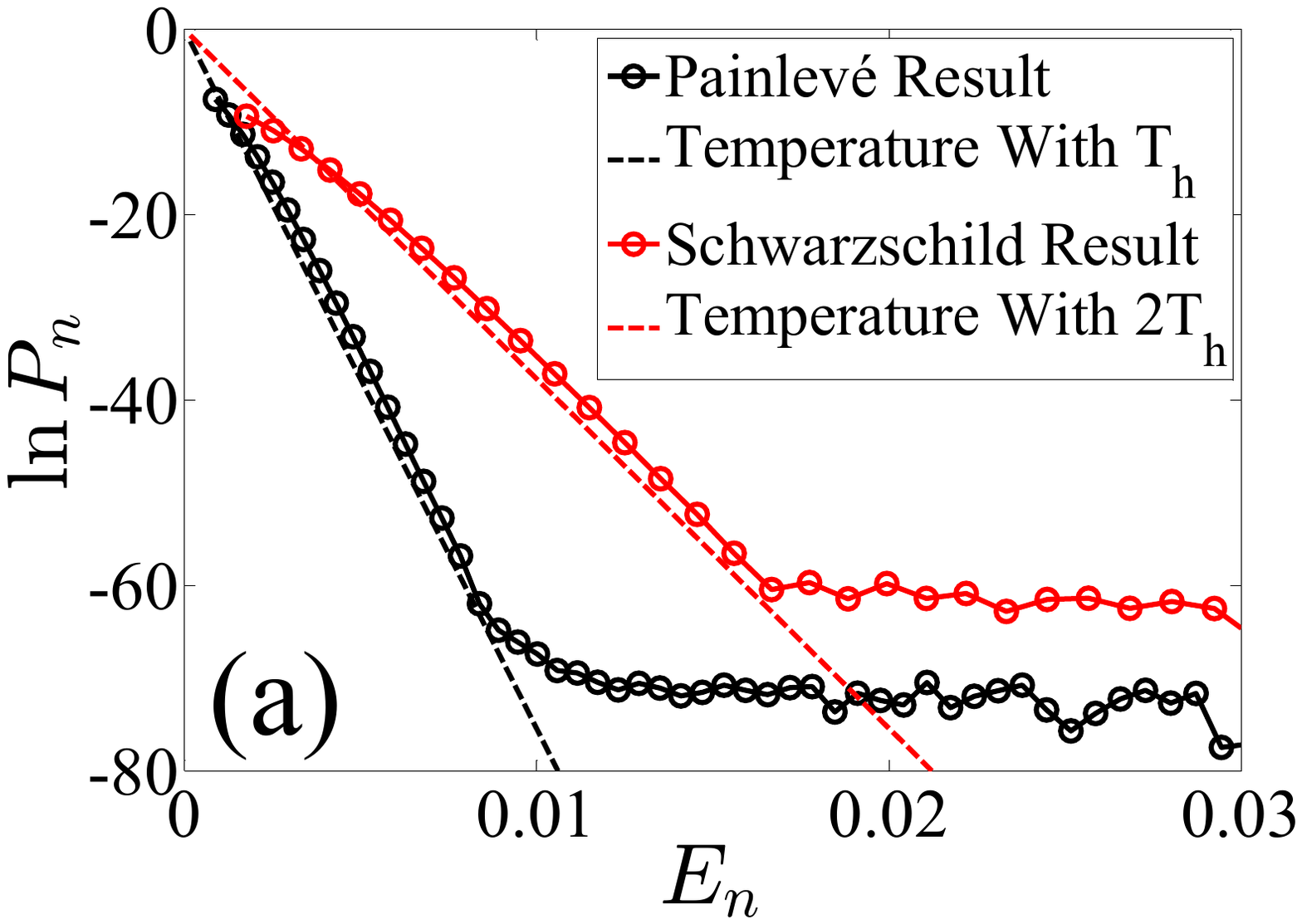}} 
            \subfigure{\includegraphics[width=0.49\linewidth,height=0.36\linewidth]{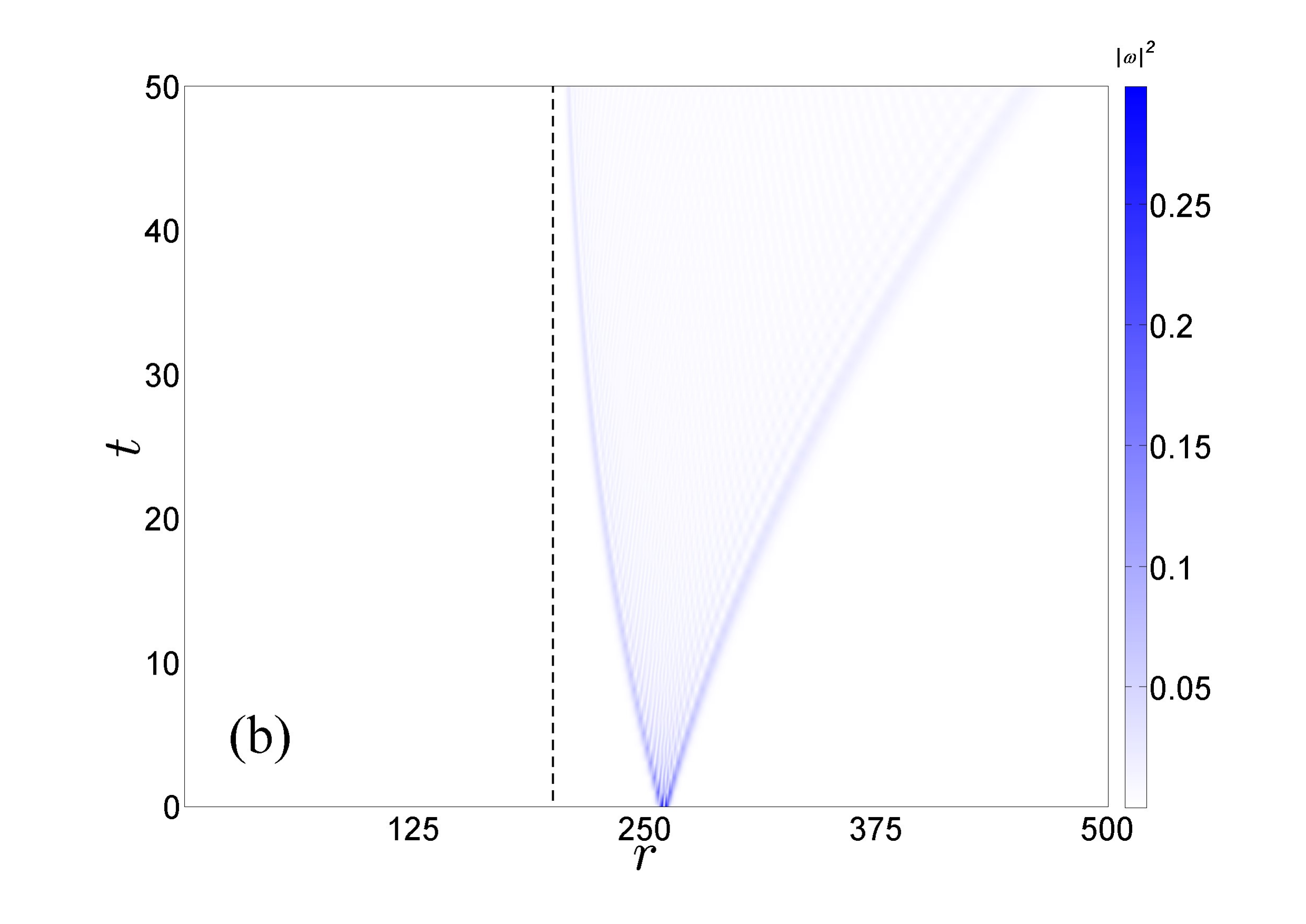}}
            }
            \caption{Numerical results. a) The Hawking Radiation Tunneling Rate. The red circle line represents the numerical results in the Schwarzschild metric, indicating the probability of detecting a particle outside the black hole, while the red dashed line represents the corresponding blackbody spectrum with equivalent temperature $2T_h$. The black circle line represents the numerical results in the Painlevé metric, and the black dashed line represents the counterpart blackbody spectrum with $T_h$.b) The dynamic evolution of the massless particle. The dashed line indicates the location of the black hole event horizon at $r=200$.}\label{fig_1a}
        \end{figure}
    Owing to the coordinate singularity present at the event horizon, this metric does not possess appropriate physical significance at that point. Consequently, it is necessary to employ two distinct vielbeins to comprehensively describe both sides. For $r>r_s$, we choose the vielbein
        \begin{equation}\label{eq_ve1} 
            e^{\mu}_{\;\; a}=
             \begin{pmatrix}
                \frac{1}{\sqrt{f(r)}} &-\sqrt{\frac{2}{f(r)}}  \\
                \sqrt{2 f(r)} & -\sqrt{f(r)} 
            \end{pmatrix} ,
        \end{equation}
    and 
        \begin{equation} \label{eq_ve2} 
            e^{\mu}_{\;\; a}=
             \begin{pmatrix}
                \frac{1}{\sqrt{-f(r)}} &0  \\
                0 & \sqrt{-f(r)} 
            \end{pmatrix} ,
        \end{equation}
    for $r<r_s$.
    Both vielbeins yield the same massless Dirac equation, which can be rewritten as
    \begin{eqnarray}
        &\partial_t \phi=-f(r)\partial_r \phi +V(r)\phi, \label{eq_14} \\ 
        &V(r)=-\frac{f^\prime(r)}{4}. 
    \end{eqnarray}
    Although the vielbeins (\ref{eq_ve1}) and (\ref{eq_ve2}) are not continuous at the horizon due to the coordinate singularity, it is noteworthy that the integral term in the transformation coefficient $Q(r)$ in Eq.(\ref{eq_tran}) is a continuous and non-divergent function, except at the physical singularity. Subsequently, following the variable substitution method mentioned earlier, the Dirac equation can be transformed into
        \begin{equation}
            \partial_t\omega=-\frac{1}{2}[f(r)\partial_r\omega+\partial_r(f(r)\omega)], \label{eq_16}
        \end{equation}
    which has been shown to be transformable into a site-dependent isotropic XY model. This model can be simulated on experimental platforms such as superconducting chips or ion traps.
    \begin{figure}[t]
        \centering{\subfigure{\includegraphics[width=0.49\linewidth,height=0.36\linewidth]{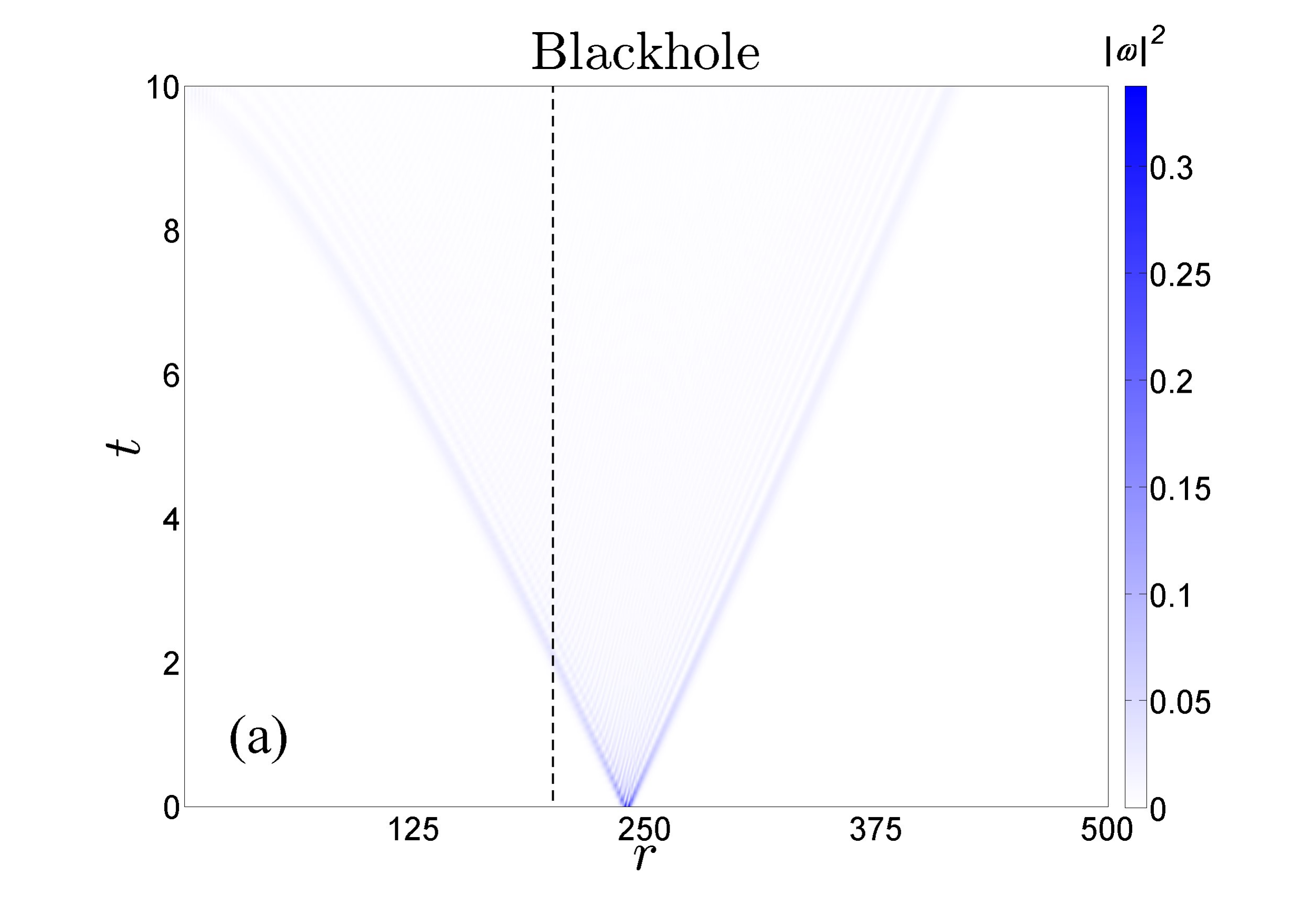}} 
        \subfigure{\includegraphics[width=0.49\linewidth,height=0.36\linewidth]{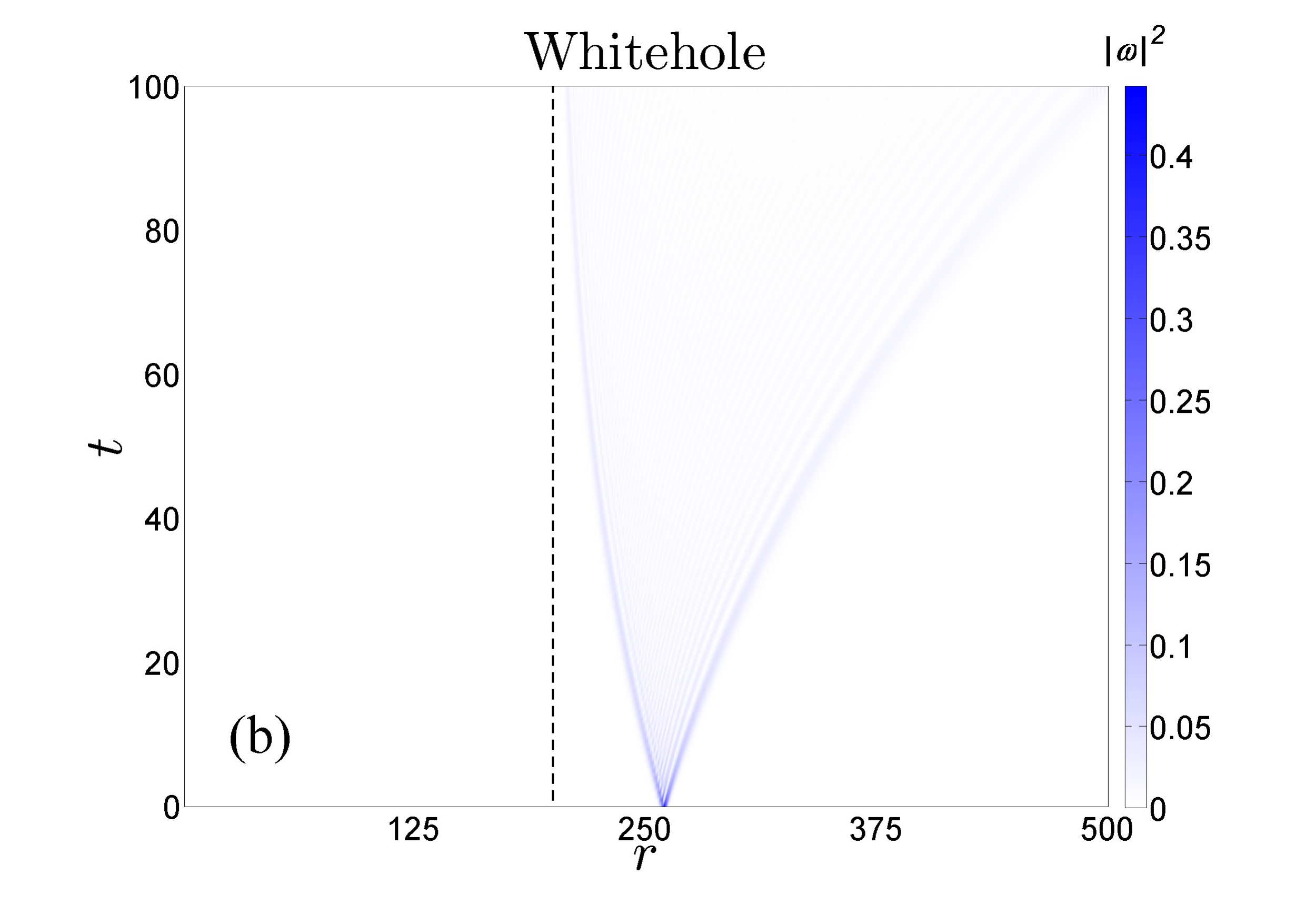}}
        }
        \caption{Numerical Simulation in Painlevé metric. a) The wave packet located outside (blue) the black hole successfully passes through the event horizon, whereas the inner segment (red) remains confined. The plus and minus signs on the color bar are employed to differentiate between these two distinct wave packets. b) The dynamics evolution of the massless particle, the dashed line indicates the locations of the black hole event horizon}\label{${2a:${fig_2a}}}
    \end{figure}
    We numerically calculated the tunneling rate in this Schwarzschild coordinate system, which is one method to study Hawking radiation \cite{damour1976black}. Although in this coordinate system the incident wave is not complete, and thus the tunneling rate cannot fully describe Hawking radiation, we can compute that the tunneling rate of the outgoing wave still satisfies the blackbody spectrum with an effective temperature of $2T_h$, where $T_h$ represents Hawking radiation (Appendix \ref{Appendix}). 
    
    The numerical results are shown in Fig.\ref{fig_1a}-(a), where the red dashed line represents the blackbody spectrum with an effective temperature of $2T_h$ and the numerical results are indicated by red circles. We find that the numerical results (red circles) are in excellent agreement with the theoretical calculations (red dashed line).
    In Fig.\ref{fig_1a}-(b), we demonstrate the evolution of an external excited particle. The black dashed line represents the position of the event horizon. As expected, as the trajectory of the massless Dirac particle represents the light cone. As approaching the horizon, its velocity continuously decreases due to the influence of the redshift factor, and it ultimately cannot reach the horizon, which is consistent with the behavior observed by an observer at infinity. Since at $a=0$, the two universes are causally disconnected but exhibit similar signal propagation behavior, we simulate only one side of the universe here.
    
    In the course of researching Hawking radiation, the singularity of the event horizon in Schwarzschild coordinates might bring about difficulties in comprehension and calculation. To achieve a clearer understanding of Hawking radiation, coordinate systems without coordinate singularities can be selected for simulation, and this is supported by the mapping method we have proposed, which supports simulations in different coordinate systems.

\subsubsection{black hole in Painlevé coordinates \texorpdfstring{$\left\{t_s,r\right\}$}{}}
        
Owing to the multi-coordinate system applicability of this mapping method, we can introduce a new coordinate system to address the issue of coordinate singularity, facilitating our study of the complete Hawking radiation effect and the process of particles crossing the event horizon. Here, we employ a new coordinate system introduced by Painlevé \cite{painleve1921comptes}. By transforming the time coordinate and imposing a condition that renders the spacetime slice Euclidean \cite{liuliao2008Nature}, we can eliminate the coordinate singularity and obtain the Painlevé metric, which is formulated as
        \begin{equation}
           \mathrm{d}s^2=-\left[1-h(r)\right]\mathrm{d}\tilde{t}^2\pm 2\sqrt{h(r)}\mathrm{d}\tilde{t}\mathrm{d}r+\mathrm{d}r^2,
        \end{equation}
    where $h(r)=\frac{r_s} {r}$ and $\tilde{t}$ is time coordinate in Painlevé coordinate system. The choice of sign in second term of the metric differentiates between distinct scenarios: the plus sign pertains to the black hole case, while the minus sign corresponds to the white hole case. In the context of a black hole, choosing the vielbein as
    \begin{equation}
        e^{\mu}_{\;\; a}=
         \begin{pmatrix}
            1 &-\sqrt{2}  \\
            -[\sqrt{2}+\sqrt{h(r)}] & [1+\sqrt{2h(r)}]  
        \end{pmatrix},
    \end{equation}
    then the Dirac equation can be written as
    \begin{equation}
        \partial_{\tilde{t}} \phi=[1+\sqrt{h(r)}]\partial_r \phi+\frac{h^\prime(r)}{4\sqrt{h(r)}}\phi.
    \end{equation}
    Through the variable transformation above and the Dirac equation can be transformed into
    \begin{equation}
        \partial_{\tilde{t}}\omega=\frac{1}{2}\left[\left(1+\sqrt{h(r)}\right)\partial_r\omega+\partial_r\left((1+\sqrt{h(r)}\right)\omega) \right].
    \end{equation}
    This equation clearly describes the incoming waves, showing that it is possible to experimentally simulate the behavior of particles falling into black holes. A similar method can be used to simulate particles moving towards a white hole. Choosing the vielbein
    \begin{equation}
        e^{\mu}_{\;\; a}=
         \begin{pmatrix}
            1 &-\sqrt{2}  \\
            -[\sqrt{2}-\sqrt{h(r)}] & [1-\sqrt{2h(r)}]  
        \end{pmatrix}. \label{vb_wh}
    \end{equation}
    Subsequently, the Dirac equation for the incoming wave can be formulated as
    \begin{equation}
        \partial_{\tilde{t}}\omega=\frac{1}{2}\left[\left(1-\sqrt{h(r)}\right)\partial_r\omega+\partial_r\left(1-\sqrt{h(r)}\right)\omega \right].
    \end{equation}
    which is just the outgoing wave equation in the black hole case as $\tilde{t}\rightarrow -\tilde{t}$. 
    
    In fact, the Painlevé metric eliminates the singularity at the event horizon, making the incoming waves complete. Therefore, we can also calculate the tunneling rate of Hawking radiation by constructing the outgoing wave equation for black holes or by using the time-reversed method (tunneling rate for incoming white holes) \cite{schutzhold2002gravity,rousseaux2008observation,weinfurtner2013classical}. Absolutely, the results of these equations are consistent. 
    
    As depicted in Fig.\ref{fig_1a}-(a), the numerical results (black circles) align with the blackbody spectrum (black dashed line), indicating that the probability $P_n$ is proportional to  $e^{-E_n/T_h}$, where $E_n$ is the energy of the particle.
    In Fig.\ref{${2a:${fig_2a}}}-(a), we illustrate the process of an external particle crossing the event horizon into a black hole and falling into the singularity within a finite time in Painlevé coordinates. It is clear that due to the elimination of the coordinate singularity at the horizon, the particle does not experience any special ``sensation" when crossing the event horizon. In Fig.\ref{${2a:${fig_2a}}}-(b), we depict the process of a particle approaching a white hole, where it experiences a significant ``repulsion", making it impossible to enter the white hole through classical process. These simulation results are in excellent agreement with the predictions of general relativity. 
    
\subsection{Regular Black Hole and Wormhole with \texorpdfstring{$a\neq 0$}{}}\label{Sec_IIIB}
    The Simpson spacetime metric exhibits a transition from a Schwarzschild black hole to a regular black hole and a wormhole as the parameter $a$ varies. Specifically, when $a\neq0$, the spacetime corresponds to a regular black hole, while for $a>r_s$, it transforms into a wormhole configuration \cite{1988Wormholes,1988WormholesII}. 
    
    We have numerically investigated the evolution of particles under different values of the parameter $a$, as depicted in Fig.\ref{${3a:${fig_3a}}}(a)-(f). The black dashed line in these figures signify the value of the regular black hole's event horizon, and the parameter $r_s$ delineates the event horizon of the Schwarzschild metric when the parameter $a=0$. Two different universes are connected at $r=0$. As illustrated, particles are unable to traverse from one universe to the other without the presence of a wormhole. 
    For $0<a<r_s$, the spacetime is a regular black hole devoid of singularities at $r=0$, and at $a=r_s$, the effective horizon shrinks to the origin of coordinates, forming a special configuration with an extremal null throat. In Fig.\ref{${3a:${fig_3a}}}(a)-(c), it is clearly demonstrated that the massless Dirac particle evolves, like the case in Schwarzschild black hole, approaching their horizon but never crossing over to the other universe. Conversely, for $a>r_s$, the horizon is disappeared and the spacetime turn into a canonical traversable wormhole geometry. In Fig.\ref{${3a:${fig_3a}}}-(d), we numerically simulate the progression of the wormhole traversal. Initially, two wave particles in each universe and then traverse the wormhole (with the throat of the wormhole located at the origin of coordinates) to the other universe as time advances. 
    In Fig.\ref{${3a:${fig_3a}}}(e)-(f), as the parameter $a$ increases, the behavior of particles traversing the wormhole gradually approaches that of flat spacetime.
    
    \begin{figure}[t]
        \centering{\subfigure{\includegraphics[width=0.49\linewidth,height=0.36\linewidth]{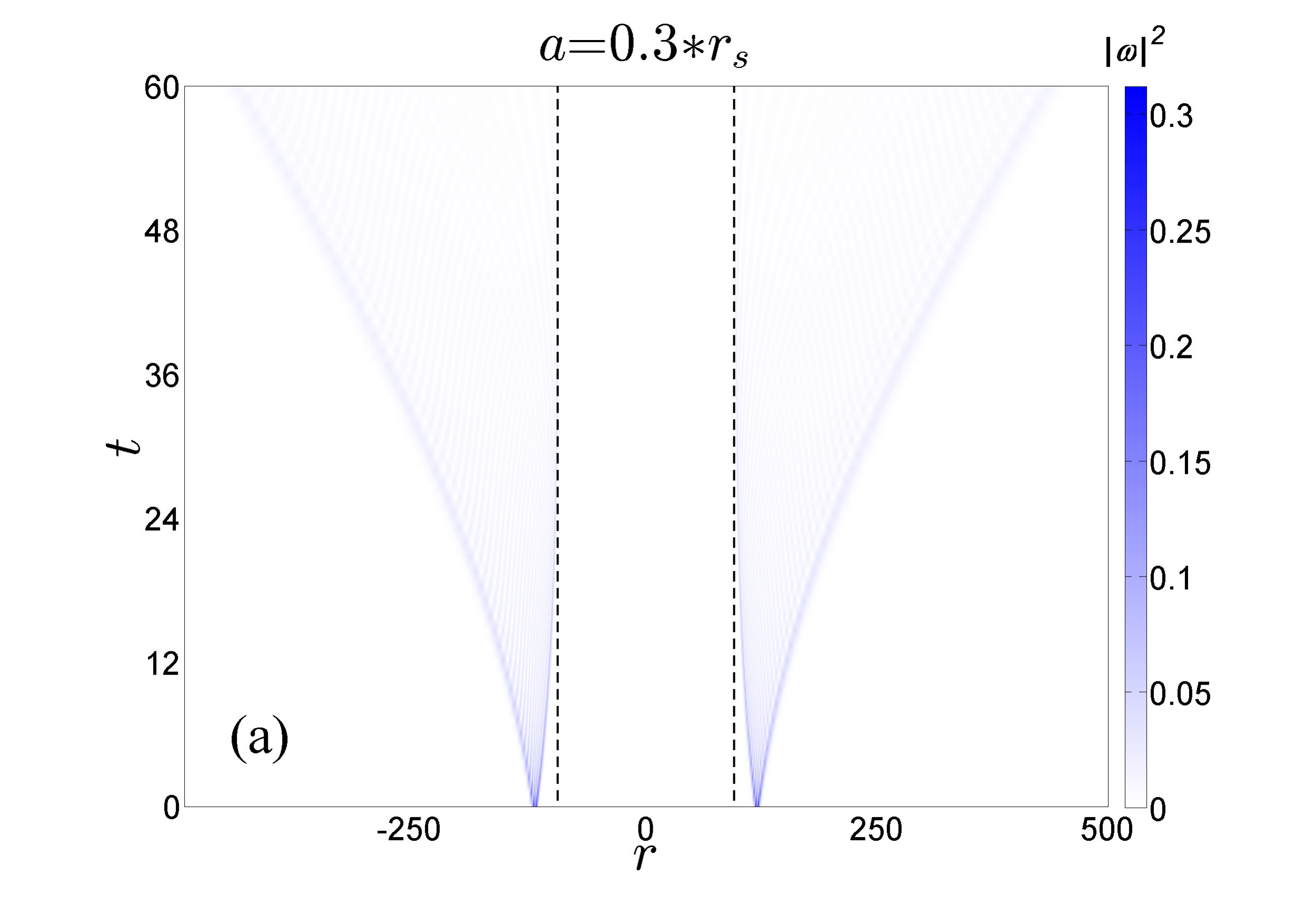}} 
        \subfigure{\includegraphics[width=0.49\linewidth,height=0.36\linewidth]{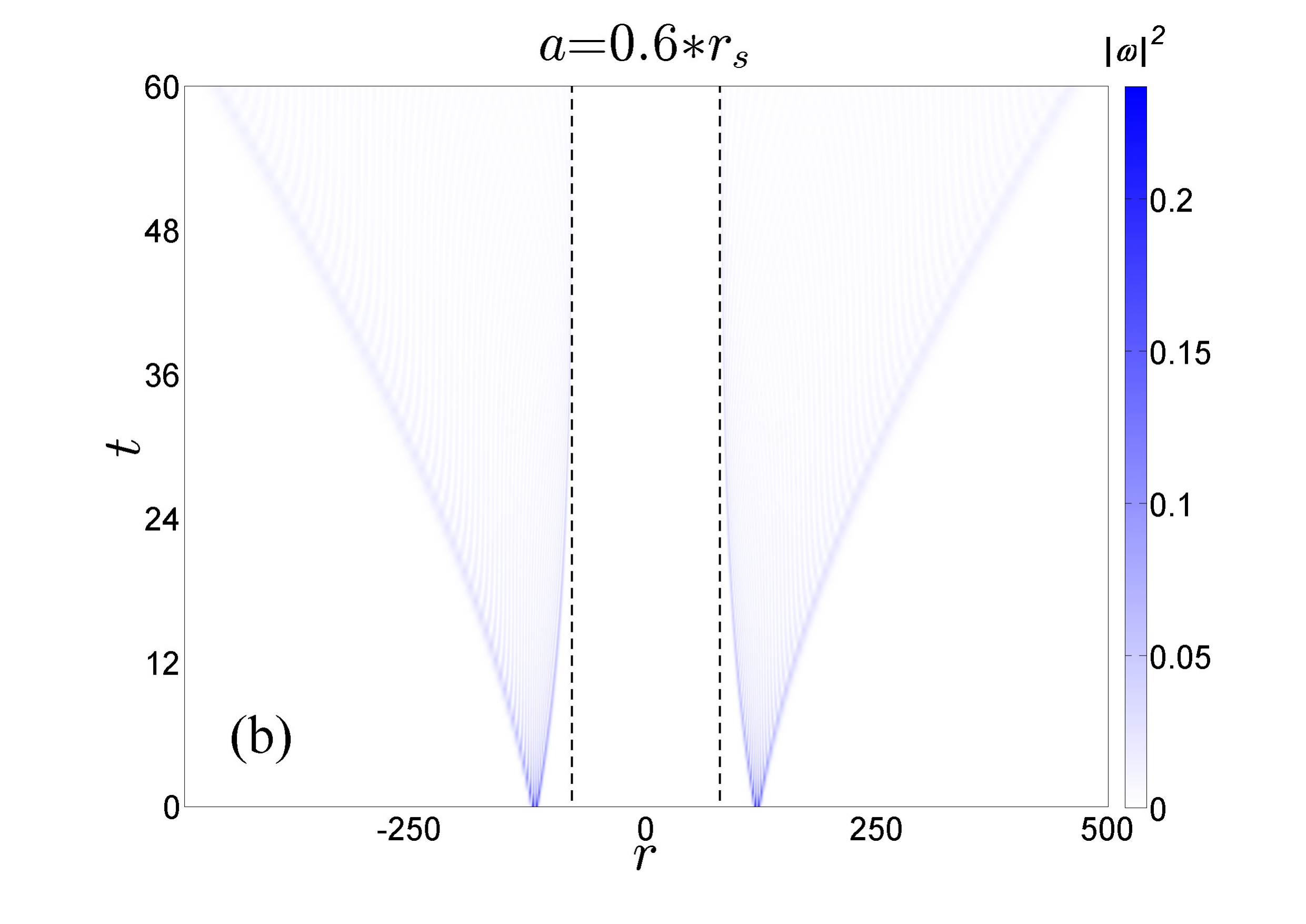}}
        \subfigure{\includegraphics[width=0.49\linewidth,height=0.36\linewidth]{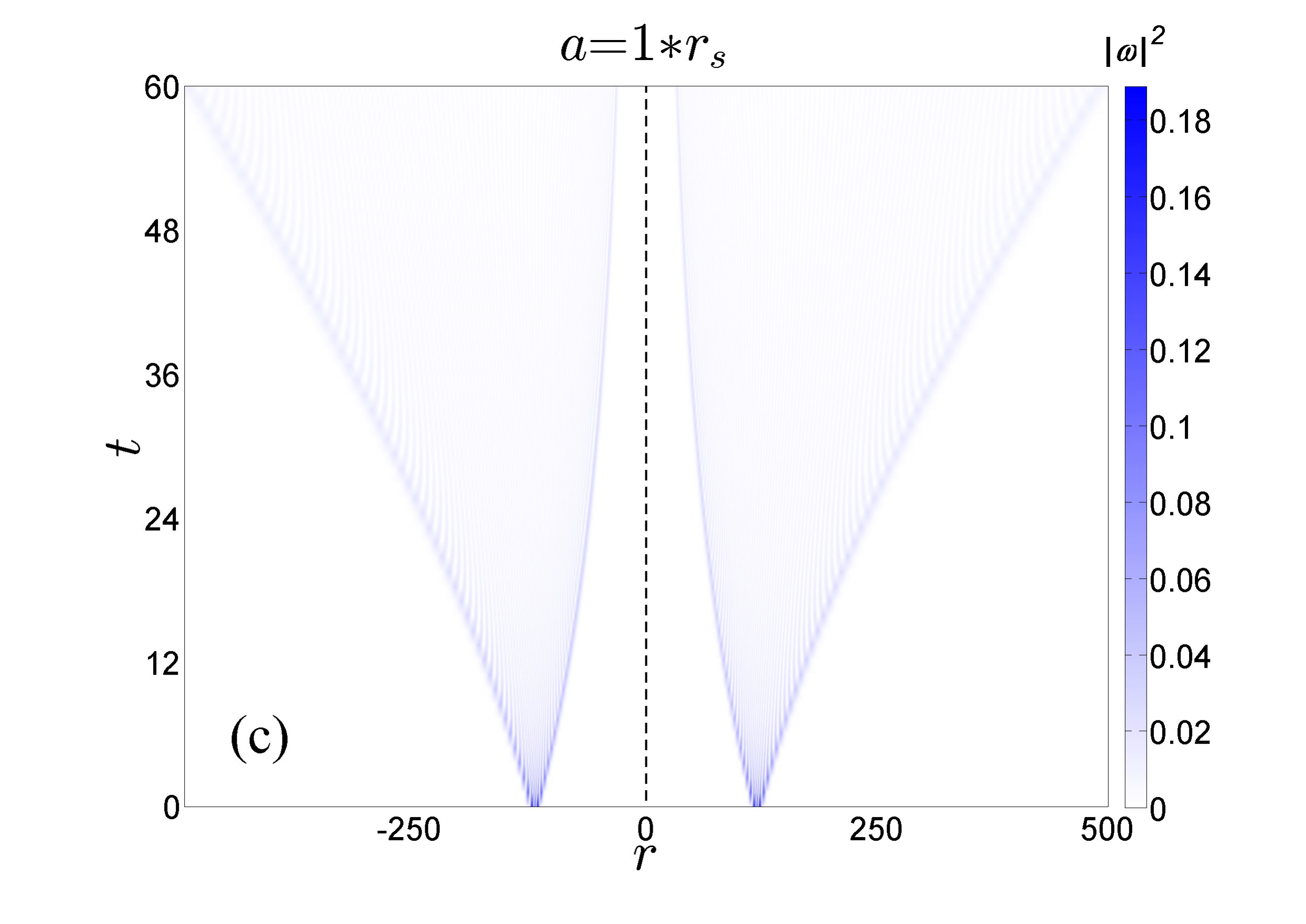}}
        \subfigure{\includegraphics[width=0.49\linewidth,height=0.36\linewidth]{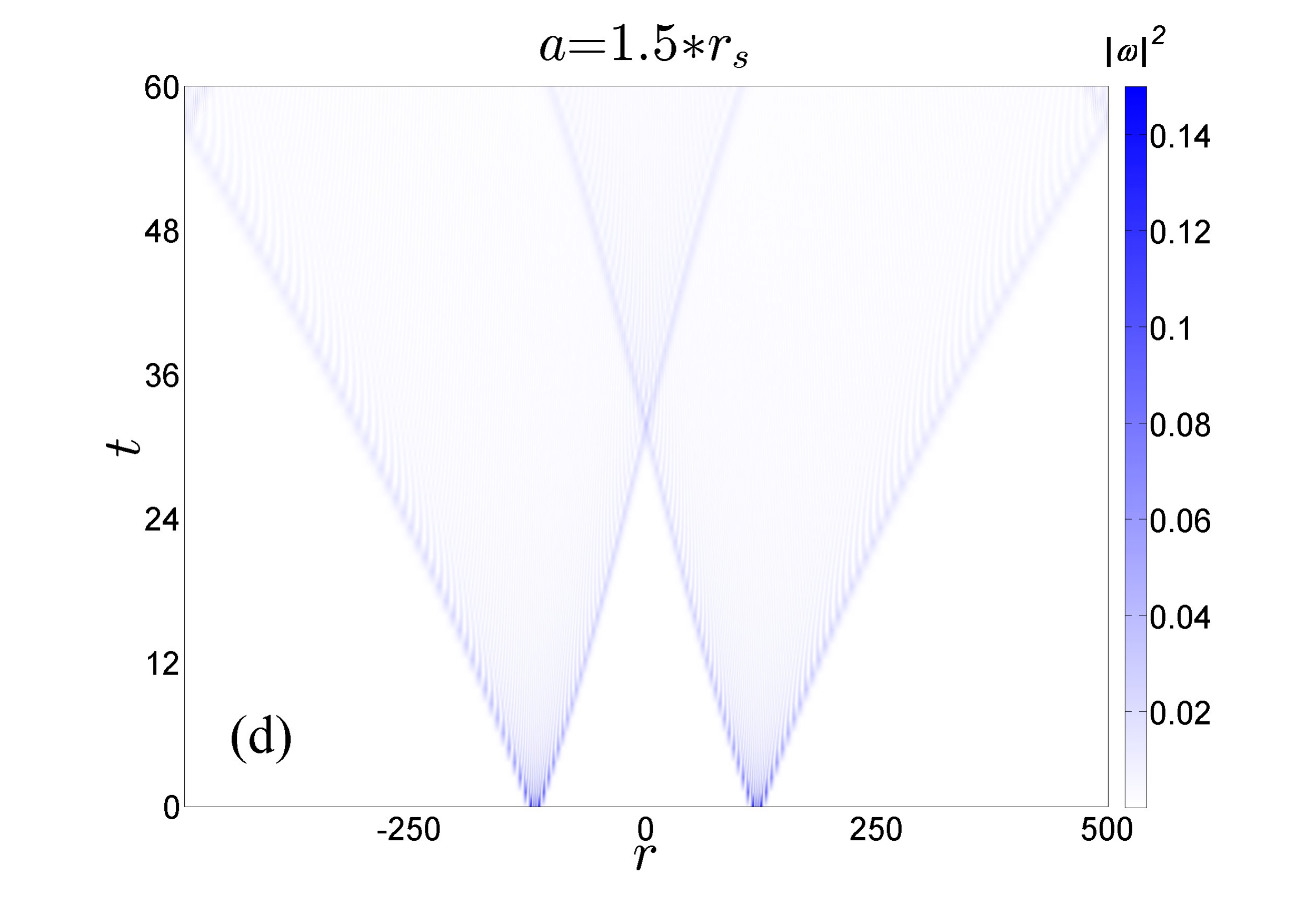}}
        \subfigure{\includegraphics[width=0.49\linewidth,height=0.36\linewidth]{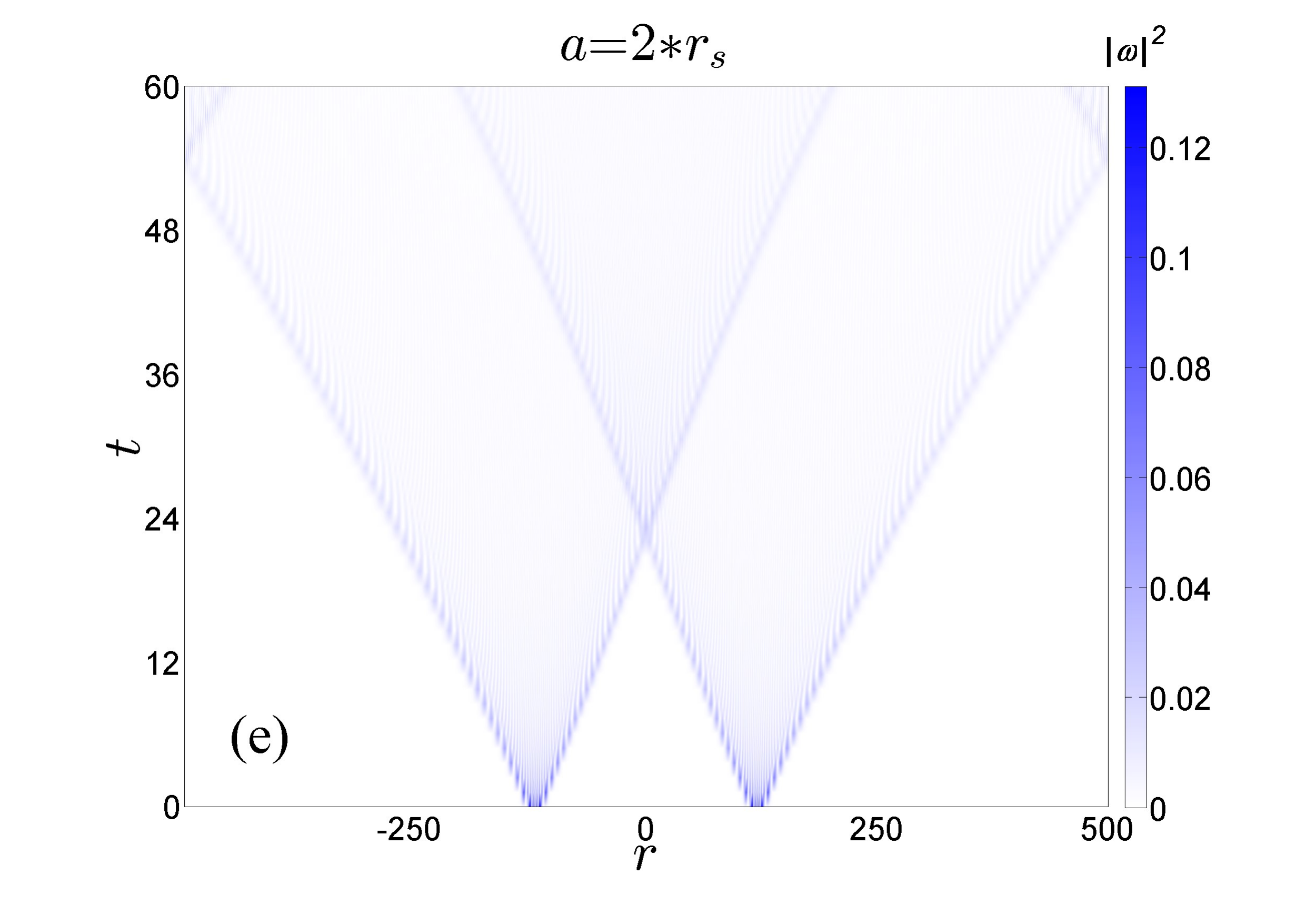}}
        \subfigure{\includegraphics[width=0.49\linewidth,height=0.36\linewidth]{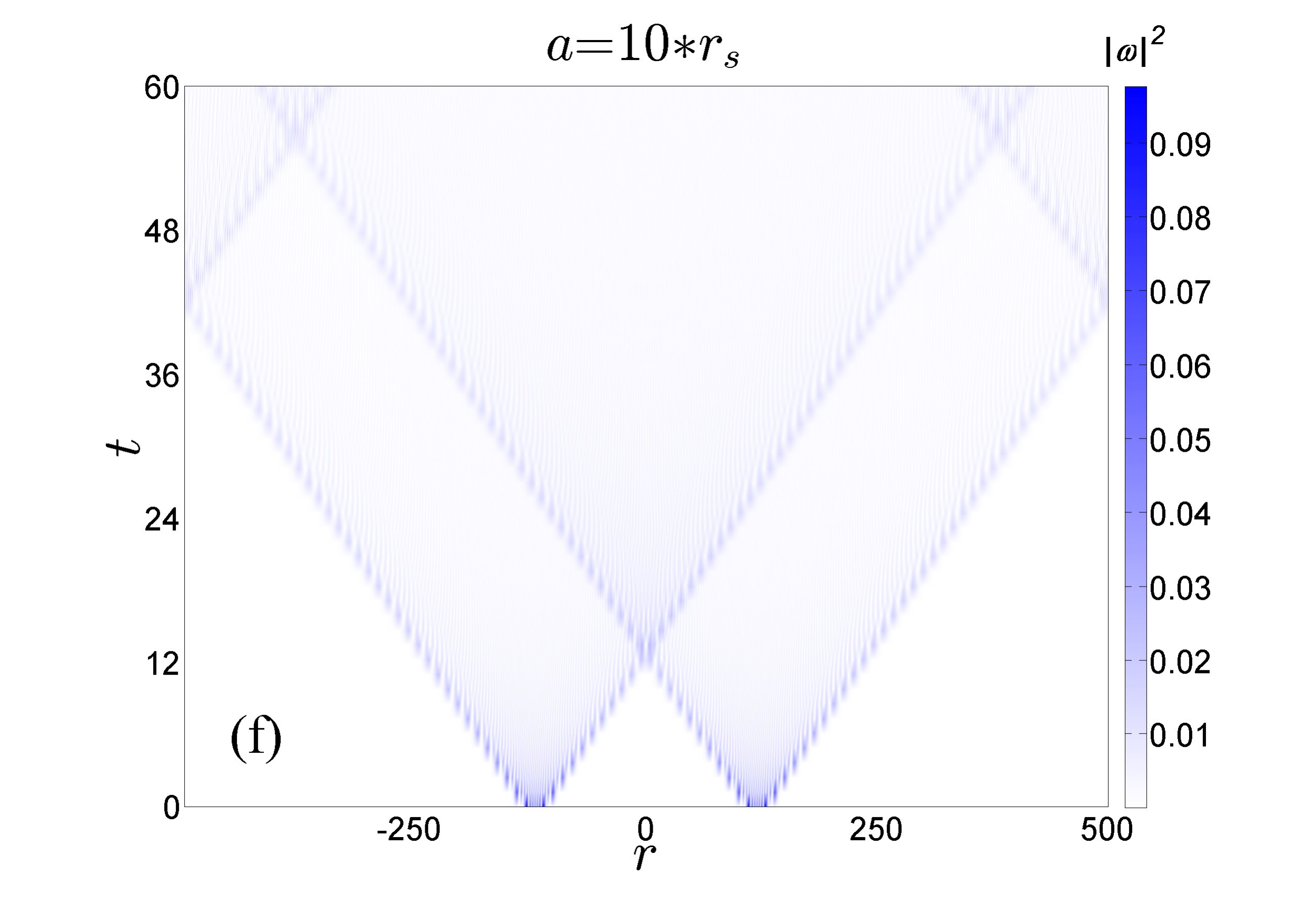}}
        }
        \caption{Numerical simulation of regular black hole and wormhole. a) - b) Regular black hole with $0<a<r_s$. c) $a=r_s$. A critical geometry with an extremal null throat at $r=0$. d)-f) A canonical traversable wormhole geometry with $a>r_s$. The dashed line indicates the different locations of equivalent event horizon. The boundary reflections observed in the figure are due to the application of Dirichlet boundary conditions.} \label{${3a:${fig_3a}}}
    \end{figure}
    Actrually, this coordinate ``$r$" in Simpson metric represent proper radial distance ``$l$ " as described in \cite{lobo2020dynamic,1988WormholesII}, rather than the radial coordinate in spherical coordinates ``$R$'', where $R\geq 0$. We can make a coordinate transformation like: $R=\sqrt{r^2+a^2}$. Then the Simpson metric can be written as 
        \begin{equation}
            \mathrm{d}s^2=-(1-\frac{r_s}{R})\mathrm{d}t^2+\frac{1}{\left(1-\frac{a^2}{R^2}\right)\left(1-\frac{r_s}{R}\right)}\mathrm{d}R^2,
        \end{equation}
    where $R\geq a$. It is explicit that this metric reduces to the Schwarzschild case as $a=0$. The parameter ``$a$" determines the position of the wormhole's throat. In this metric, it is evident that the position of the wormhole's throat moves outward as the parameter ``$a$" increases. Ultimately, as the throat moves beyond the horizon, the wave packets can traverse it to the other universe. As $a$ becomes much greater than the Schwarzschild radius which means $``R \gg r_s"$, the entire spacetime metric transforms into
    \begin{equation}
        \mathrm{d}s^2=-\mathrm{d}t^2+\frac{1}{\left(1-\frac{a^2}{R^2}\right)}\mathrm{d}R^2,
    \end{equation}
    or
        \begin{equation}
            \mathrm{d}s^2=-\mathrm{d}t^2+\mathrm{d}r^2.
        \end{equation}
    This precisely corresponds to the wormhole metric provided by Thorne et al \cite{1988Wormholes,1988WormholesII}. 

\section{Conclusion}\label{Sec_IV}
The applicability of the method to various observers is noteworthy, as it facilitates the comprehension of intricate theories from multiple perspectives and broadens the approach to studying exotic effects in curved spacetime.
By employing this method, we demonstrated the mapping of the dynamics effect of the 1+1D massless Dirac equation onto the XY model or the Hubbard model, which can be experimentally realized using platforms such as superconducting circuits or ion traps. We subsequently numerically performed simulations to illustrate infinite redshift effect and Hawking radiation tunneling rate. Furthermore, we have simulated the dynamic evolution of particles in Simpson spacetime with different parameter values of ``$a$" and the process by which massless Dirac particles traverse through wormholes. This method aids in exploring quantum effects in curved spacetime from multiple perspectives. Future work may extend these simulations to more complex spacetimes, further investigating the impact of curved spacetime on quantum fields and quantum resources.
\acknowledgments
    This work was supported by the National Natural Science Foundation of China under Grants  No. 12475051,  No. 12375051,  No. 12122504, and No. 12421005; the science and technology innovation Program of Hunan Province under grant No. 2024RC1050; the Natural Science Foundation of Hunan Province under grant No. 2023JJ30384; and the innovative research group of Hunan Province under Grant No. 2024JJ1006.

\renewcommand{\appendixname}{Appendix}

\section{Appendix}
\appendix
\section*{Hawking radiation in Schwarzschild coordinates}\label{Appendix}
        In this Appendix, we show prove process of the Hawking radiation in Schwarzschild coordinates. Choose different vielbein, we can find the equation (\ref{eq_14}) can be written as (we can eliminat the potential energy part by a transformation $\phi=Q(r)\varphi$, and $|\frac{\phi_{out2}}{\phi_{out1}}|=|\frac{\varphi_{out2}}{\varphi_{out1}}|$):
        \begin{equation}
            \partial_t \varphi=- f(r)\partial_r \varphi.
        \end{equation}
        This equation characterizes the behavior of outgoing waves, and its solution can be expressed as:
        \begin{equation}
            \varphi_{out2}^{E_n}= e^{-iE_n (t-r_*)}.
        \end{equation}
        Here, $E_n$ is the energy of the massless Dirac particle and $r_*=\int{\frac{1}{f(r)}dr}$ is tortoise coordinate. This solution is not continuous at the horizon. Then we can repeat the traditional process \cite{damour1976black,yang2020simulating}, applying $r_*=r+r_s\;ln|\frac{r-r_s}{r_s}|$. This allows us to rewrite the outgoing waves as:
        \begin{equation}
            \varphi_{out2}^{E_n} =e^{-iE_n {t-r}}(|\frac{r-r_s}{r_s}|)^{iE_n r_s}.
        \end{equation}
        This formulation differs by a factor of two from \cite{damour1976black}.  We repeat the process:  complex extension and use Heaviside function Y($r_s-r$) to describes the inner rigion,
        and then we can find that the solution in the inner region will have an additional factor $\Gamma$.
        And we can also find that 
        \begin{equation}
            \frac{|\varphi_{out2}|^2}{|\varphi_{out1}|^2}=\Gamma^2=e^{-\frac{E_n}{2T_h}}.
        \end{equation}
        It should be emphasized that this result does not imply a variation in the black hole's temperature, as the incoming waves are discontinuous at the horizon. Although the coordinate system employed is not the most suitable for describing Hawking radiation, it is noteworthy that the outgoing waves still exhibit a blackbody spectrum. Indeed, the incoming waves are complete in the Painlevé metric, and we can determine that the tunneling rate represents the Hawking radiation.
        

\begin{thebibliography}{42}%
\makeatletter
\providecommand \@ifxundefined [1]{%
 \@ifx{#1\undefined}
}%
\providecommand \@ifnum [1]{%
 \ifnum #1\expandafter \@firstoftwo
 \else \expandafter \@secondoftwo
 \fi
}%
\providecommand \@ifx [1]{%
 \ifx #1\expandafter \@firstoftwo
 \else \expandafter \@secondoftwo
 \fi
}%
\providecommand \natexlab [1]{#1}%
\providecommand \enquote  [1]{``#1''}%
\providecommand \bibnamefont  [1]{#1}%
\providecommand \bibfnamefont [1]{#1}%
\providecommand \citenamefont [1]{#1}%
\providecommand \href@noop [0]{\@secondoftwo}%
\providecommand \href [0]{\begingroup \@sanitize@url \@href}%
\providecommand \@href[1]{\@@startlink{#1}\@@href}%
\providecommand \@@href[1]{\endgroup#1\@@endlink}%
\providecommand \@sanitize@url [0]{\catcode `\\12\catcode `\$12\catcode `\&12\catcode `\#12\catcode `\^12\catcode `\_12\catcode `\%12\relax}%
\providecommand \@@startlink[1]{}%
\providecommand \@@endlink[0]{}%
\providecommand \url  [0]{\begingroup\@sanitize@url \@url }%
\providecommand \@url [1]{\endgroup\@href {#1}{\urlprefix }}%
\providecommand \urlprefix  [0]{URL }%
\providecommand \Eprint [0]{\href }%
\providecommand \doibase [0]{https://doi.org/}%
\providecommand \selectlanguage [0]{\@gobble}%
\providecommand \bibinfo  [0]{\@secondoftwo}%
\providecommand \bibfield  [0]{\@secondoftwo}%
\providecommand \translation [1]{[#1]}%
\providecommand \BibitemOpen [0]{}%
\providecommand \bibitemStop [0]{}%
\providecommand \bibitemNoStop [0]{.\EOS\space}%
\providecommand \EOS [0]{\spacefactor3000\relax}%
\providecommand \BibitemShut  [1]{\csname bibitem#1\endcsname}%
\let\auto@bib@innerbib\@empty
\bibitem [{\citenamefont {Casimir}(1948)}]{casimir1948attraction}%
  \BibitemOpen
  \bibfield  {author} {\bibinfo {author} {\bibfnamefont {H.~B.}\ \bibnamefont {Casimir}},\ }\bibfield  {title} {\bibinfo {title} {On the attraction between two perfectly conducting plates},\ }in\ \href@noop {} {\emph {\bibinfo {booktitle} {Proc. Kon. Ned. Akad. Wet.}}},\ Vol.~\bibinfo {volume} {51}\ (\bibinfo {year} {1948})\ p.\ \bibinfo {pages} {793}\BibitemShut {NoStop}%
\bibitem [{\citenamefont {Parker}(1968)}]{parker1968particle}%
  \BibitemOpen
  \bibfield  {author} {\bibinfo {author} {\bibfnamefont {L.}~\bibnamefont {Parker}},\ }\bibfield  {title} {\bibinfo {title} {Particle creation in expanding universes},\ }\href@noop {} {\bibfield  {journal} {\bibinfo  {journal} {Physical Review Letters}\ }\textbf {\bibinfo {volume} {21}},\ \bibinfo {pages} {562} (\bibinfo {year} {1968})}\BibitemShut {NoStop}%
\bibitem [{\citenamefont {Hawking}(1974)}]{hawking1974black}%
  \BibitemOpen
  \bibfield  {author} {\bibinfo {author} {\bibfnamefont {S.~W.}\ \bibnamefont {Hawking}},\ }\bibfield  {title} {\bibinfo {title} {Black hole explosions?},\ }\href@noop {} {\bibfield  {journal} {\bibinfo  {journal} {Nature}\ }\textbf {\bibinfo {volume} {248}},\ \bibinfo {pages} {30} (\bibinfo {year} {1974})}\BibitemShut {NoStop}%
\bibitem [{\citenamefont {Unruh}(1976)}]{unruh1976notes}%
  \BibitemOpen
  \bibfield  {author} {\bibinfo {author} {\bibfnamefont {W.~G.}\ \bibnamefont {Unruh}},\ }\bibfield  {title} {\bibinfo {title} {Notes on black-hole evaporation},\ }\href@noop {} {\bibfield  {journal} {\bibinfo  {journal} {Physical Review D}\ }\textbf {\bibinfo {volume} {14}},\ \bibinfo {pages} {870} (\bibinfo {year} {1976})}\BibitemShut {NoStop}%
\bibitem [{\citenamefont {Unruh}(1981)}]{unruh1981experimental}%
  \BibitemOpen
  \bibfield  {author} {\bibinfo {author} {\bibfnamefont {W.~G.}\ \bibnamefont {Unruh}},\ }\bibfield  {title} {\bibinfo {title} {Experimental black-hole evaporation?},\ }\href@noop {} {\bibfield  {journal} {\bibinfo  {journal} {Physical Review Letters}\ }\textbf {\bibinfo {volume} {46}},\ \bibinfo {pages} {1351} (\bibinfo {year} {1981})}\BibitemShut {NoStop}%
\bibitem [{\citenamefont {Rousseaux}\ \emph {et~al.}(2008)\citenamefont {Rousseaux}, \citenamefont {Mathis}, \citenamefont {Ma{\"\i}ssa}, \citenamefont {Philbin},\ and\ \citenamefont {Leonhardt}}]{rousseaux2008observation}%
  \BibitemOpen
  \bibfield  {author} {\bibinfo {author} {\bibfnamefont {G.}~\bibnamefont {Rousseaux}}, \bibinfo {author} {\bibfnamefont {C.}~\bibnamefont {Mathis}}, \bibinfo {author} {\bibfnamefont {P.}~\bibnamefont {Ma{\"\i}ssa}}, \bibinfo {author} {\bibfnamefont {T.~G.}\ \bibnamefont {Philbin}},\ and\ \bibinfo {author} {\bibfnamefont {U.}~\bibnamefont {Leonhardt}},\ }\bibfield  {title} {\bibinfo {title} {Observation of negative-frequency waves in a water tank: a classical analogue to the hawking effect?},\ }\href@noop {} {\bibfield  {journal} {\bibinfo  {journal} {New Journal of Physics}\ }\textbf {\bibinfo {volume} {10}},\ \bibinfo {pages} {053015} (\bibinfo {year} {2008})}\BibitemShut {NoStop}%
\bibitem [{\citenamefont {Weinfurtner}\ \emph {et~al.}(2011)\citenamefont {Weinfurtner}, \citenamefont {Tedford}, \citenamefont {Penrice}, \citenamefont {Unruh},\ and\ \citenamefont {Lawrence}}]{weinfurtner2011measurement}%
  \BibitemOpen
  \bibfield  {author} {\bibinfo {author} {\bibfnamefont {S.}~\bibnamefont {Weinfurtner}}, \bibinfo {author} {\bibfnamefont {E.~W.}\ \bibnamefont {Tedford}}, \bibinfo {author} {\bibfnamefont {M.~C.}\ \bibnamefont {Penrice}}, \bibinfo {author} {\bibfnamefont {W.~G.}\ \bibnamefont {Unruh}},\ and\ \bibinfo {author} {\bibfnamefont {G.~A.}\ \bibnamefont {Lawrence}},\ }\bibfield  {title} {\bibinfo {title} {Measurement of stimulated hawking emission in an analogue system},\ }\href@noop {} {\bibfield  {journal} {\bibinfo  {journal} {Physical Review Letters}\ }\textbf {\bibinfo {volume} {106}},\ \bibinfo {pages} {021302} (\bibinfo {year} {2011})}\BibitemShut {NoStop}%
\bibitem [{\citenamefont {Braunstein}\ \emph {et~al.}(2023)\citenamefont {Braunstein}, \citenamefont {Faizal}, \citenamefont {Krauss}, \citenamefont {Marino},\ and\ \citenamefont {Shah}}]{braunstein2023analogue}%
  \BibitemOpen
  \bibfield  {author} {\bibinfo {author} {\bibfnamefont {S.~L.}\ \bibnamefont {Braunstein}}, \bibinfo {author} {\bibfnamefont {M.}~\bibnamefont {Faizal}}, \bibinfo {author} {\bibfnamefont {L.~M.}\ \bibnamefont {Krauss}}, \bibinfo {author} {\bibfnamefont {F.}~\bibnamefont {Marino}},\ and\ \bibinfo {author} {\bibfnamefont {N.~A.}\ \bibnamefont {Shah}},\ }\bibfield  {title} {\bibinfo {title} {Analogue simulations of quantum gravity with fluids},\ }\href@noop {} {\bibfield  {journal} {\bibinfo  {journal} {Nature Reviews Physics}\ }\textbf {\bibinfo {volume} {5}},\ \bibinfo {pages} {612} (\bibinfo {year} {2023})}\BibitemShut {NoStop}%
\bibitem [{\citenamefont {M{\"a}kinen}\ \emph {et~al.}(2023)\citenamefont {M{\"a}kinen}, \citenamefont {Autti}, \citenamefont {Heikkinen}, \citenamefont {Hosio}, \citenamefont {H{\"a}nninen}, \citenamefont {L’vov}, \citenamefont {Walmsley}, \citenamefont {Zavjalov},\ and\ \citenamefont {Eltsov}}]{makinen2023rotating}%
  \BibitemOpen
  \bibfield  {author} {\bibinfo {author} {\bibfnamefont {J.}~\bibnamefont {M{\"a}kinen}}, \bibinfo {author} {\bibfnamefont {S.}~\bibnamefont {Autti}}, \bibinfo {author} {\bibfnamefont {P.}~\bibnamefont {Heikkinen}}, \bibinfo {author} {\bibfnamefont {J.}~\bibnamefont {Hosio}}, \bibinfo {author} {\bibfnamefont {R.}~\bibnamefont {H{\"a}nninen}}, \bibinfo {author} {\bibfnamefont {V.}~\bibnamefont {L’vov}}, \bibinfo {author} {\bibfnamefont {P.}~\bibnamefont {Walmsley}}, \bibinfo {author} {\bibfnamefont {V.}~\bibnamefont {Zavjalov}},\ and\ \bibinfo {author} {\bibfnamefont {V.}~\bibnamefont {Eltsov}},\ }\bibfield  {title} {\bibinfo {title} {Rotating quantum wave turbulence},\ }\href@noop {} {\bibfield  {journal} {\bibinfo  {journal} {Nature physics}\ }\textbf {\bibinfo {volume} {19}},\ \bibinfo {pages} {898} (\bibinfo {year} {2023})}\BibitemShut {NoStop}%
\bibitem [{\citenamefont {{\v{S}}van{\v{c}}ara}\ \emph {et~al.}(2024)\citenamefont {{\v{S}}van{\v{c}}ara}, \citenamefont {Smaniotto}, \citenamefont {Solidoro}, \citenamefont {MacDonald}, \citenamefont {Patrick}, \citenamefont {Gregory}, \citenamefont {Barenghi},\ and\ \citenamefont {Weinfurtner}}]{vsvanvcara2024rotating}%
  \BibitemOpen
  \bibfield  {author} {\bibinfo {author} {\bibfnamefont {P.}~\bibnamefont {{\v{S}}van{\v{c}}ara}}, \bibinfo {author} {\bibfnamefont {P.}~\bibnamefont {Smaniotto}}, \bibinfo {author} {\bibfnamefont {L.}~\bibnamefont {Solidoro}}, \bibinfo {author} {\bibfnamefont {J.~F.}\ \bibnamefont {MacDonald}}, \bibinfo {author} {\bibfnamefont {S.}~\bibnamefont {Patrick}}, \bibinfo {author} {\bibfnamefont {R.}~\bibnamefont {Gregory}}, \bibinfo {author} {\bibfnamefont {C.~F.}\ \bibnamefont {Barenghi}},\ and\ \bibinfo {author} {\bibfnamefont {S.}~\bibnamefont {Weinfurtner}},\ }\bibfield  {title} {\bibinfo {title} {Rotating curved spacetime signatures from a giant quantum vortex},\ }\href@noop {} {\bibfield  {journal} {\bibinfo  {journal} {Nature}\ }\textbf {\bibinfo {volume} {628}},\ \bibinfo {pages} {66} (\bibinfo {year} {2024})}\BibitemShut {NoStop}%
\bibitem [{\citenamefont {Garay}\ \emph {et~al.}(2000)\citenamefont {Garay}, \citenamefont {Anglin}, \citenamefont {Cirac},\ and\ \citenamefont {Zoller}}]{garay2000sonic}%
  \BibitemOpen
  \bibfield  {author} {\bibinfo {author} {\bibfnamefont {L.~J.}\ \bibnamefont {Garay}}, \bibinfo {author} {\bibfnamefont {J.}~\bibnamefont {Anglin}}, \bibinfo {author} {\bibfnamefont {J.~I.}\ \bibnamefont {Cirac}},\ and\ \bibinfo {author} {\bibfnamefont {P.}~\bibnamefont {Zoller}},\ }\bibfield  {title} {\bibinfo {title} {Sonic analog of gravitational black holes in bose-einstein condensates},\ }\href@noop {} {\bibfield  {journal} {\bibinfo  {journal} {Physical Review Letters}\ }\textbf {\bibinfo {volume} {85}},\ \bibinfo {pages} {4643} (\bibinfo {year} {2000})}\BibitemShut {NoStop}%
\bibitem [{\citenamefont {Steinhauer}(2016)}]{steinhauer2016observation}%
  \BibitemOpen
  \bibfield  {author} {\bibinfo {author} {\bibfnamefont {J.}~\bibnamefont {Steinhauer}},\ }\bibfield  {title} {\bibinfo {title} {Observation of quantum hawking radiation and its entanglement in an analogue black hole},\ }\href@noop {} {\bibfield  {journal} {\bibinfo  {journal} {Nature Physics}\ }\textbf {\bibinfo {volume} {12}},\ \bibinfo {pages} {959} (\bibinfo {year} {2016})}\BibitemShut {NoStop}%
\bibitem [{\citenamefont {Eckel}\ \emph {et~al.}(2018)\citenamefont {Eckel}, \citenamefont {Kumar}, \citenamefont {Jacobson}, \citenamefont {Spielman},\ and\ \citenamefont {Campbell}}]{eckel2018rapidly}%
  \BibitemOpen
  \bibfield  {author} {\bibinfo {author} {\bibfnamefont {S.}~\bibnamefont {Eckel}}, \bibinfo {author} {\bibfnamefont {A.}~\bibnamefont {Kumar}}, \bibinfo {author} {\bibfnamefont {T.}~\bibnamefont {Jacobson}}, \bibinfo {author} {\bibfnamefont {I.~B.}\ \bibnamefont {Spielman}},\ and\ \bibinfo {author} {\bibfnamefont {G.~K.}\ \bibnamefont {Campbell}},\ }\bibfield  {title} {\bibinfo {title} {A rapidly expanding bose-einstein condensate: an expanding universe in the lab},\ }\href@noop {} {\bibfield  {journal} {\bibinfo  {journal} {Physical Review X}\ }\textbf {\bibinfo {volume} {8}},\ \bibinfo {pages} {021021} (\bibinfo {year} {2018})}\BibitemShut {NoStop}%
\bibitem [{\citenamefont {Tian}\ and\ \citenamefont {Du}(2021)}]{tian2021probing2}%
  \BibitemOpen
  \bibfield  {author} {\bibinfo {author} {\bibfnamefont {Z.}~\bibnamefont {Tian}}\ and\ \bibinfo {author} {\bibfnamefont {J.}~\bibnamefont {Du}},\ }\bibfield  {title} {\bibinfo {title} {Probing low-energy lorentz violation from high-energy modified dispersion in dipolar bose-einstein condensates},\ }\href@noop {} {\bibfield  {journal} {\bibinfo  {journal} {Physical Review D}\ }\textbf {\bibinfo {volume} {103}},\ \bibinfo {pages} {085014} (\bibinfo {year} {2021})}\BibitemShut {NoStop}%
\bibitem [{\citenamefont {Tian}\ \emph {et~al.}(2022)\citenamefont {Tian}, \citenamefont {Wu}, \citenamefont {Zhang}, \citenamefont {Jing},\ and\ \citenamefont {Du}}]{tian2022probing}%
  \BibitemOpen
  \bibfield  {author} {\bibinfo {author} {\bibfnamefont {Z.}~\bibnamefont {Tian}}, \bibinfo {author} {\bibfnamefont {L.}~\bibnamefont {Wu}}, \bibinfo {author} {\bibfnamefont {L.}~\bibnamefont {Zhang}}, \bibinfo {author} {\bibfnamefont {J.}~\bibnamefont {Jing}},\ and\ \bibinfo {author} {\bibfnamefont {J.}~\bibnamefont {Du}},\ }\bibfield  {title} {\bibinfo {title} {Probing lorentz-invariance-violation-induced nonthermal unruh effect in quasi-two-dimensional dipolar condensates},\ }\href@noop {} {\bibfield  {journal} {\bibinfo  {journal} {Physical Review D}\ }\textbf {\bibinfo {volume} {106}},\ \bibinfo {pages} {L061701} (\bibinfo {year} {2022})}\BibitemShut {NoStop}%
\bibitem [{\citenamefont {Blencowe}\ and\ \citenamefont {Wang}(2020)}]{blencowe2020analogue}%
  \BibitemOpen
  \bibfield  {author} {\bibinfo {author} {\bibfnamefont {M.~P.}\ \bibnamefont {Blencowe}}\ and\ \bibinfo {author} {\bibfnamefont {H.}~\bibnamefont {Wang}},\ }\bibfield  {title} {\bibinfo {title} {Analogue gravity on a superconducting chip},\ }\href@noop {} {\bibfield  {journal} {\bibinfo  {journal} {Philosophical Transactions of the Royal Society A}\ }\textbf {\bibinfo {volume} {378}},\ \bibinfo {pages} {20190224} (\bibinfo {year} {2020})}\BibitemShut {NoStop}%
\bibitem [{\citenamefont {Terrones}\ and\ \citenamefont {Sab{\'\i}n}(2021)}]{terrones2021towards}%
  \BibitemOpen
  \bibfield  {author} {\bibinfo {author} {\bibfnamefont {A.}~\bibnamefont {Terrones}}\ and\ \bibinfo {author} {\bibfnamefont {C.}~\bibnamefont {Sab{\'\i}n}},\ }\bibfield  {title} {\bibinfo {title} {Towards quantum simulation of black holes in a dc-squid array},\ }\href@noop {} {\bibfield  {journal} {\bibinfo  {journal} {Universe}\ }\textbf {\bibinfo {volume} {7}},\ \bibinfo {pages} {499} (\bibinfo {year} {2021})}\BibitemShut {NoStop}%
\bibitem [{\citenamefont {Liu}\ \emph {et~al.}(2014)\citenamefont {Liu}, \citenamefont {Sheng}, \citenamefont {Zhu},\ and\ \citenamefont {Genov}}]{liu2014trapping}%
  \BibitemOpen
  \bibfield  {author} {\bibinfo {author} {\bibfnamefont {H.}~\bibnamefont {Liu}}, \bibinfo {author} {\bibfnamefont {C.}~\bibnamefont {Sheng}}, \bibinfo {author} {\bibfnamefont {S.}~\bibnamefont {Zhu}},\ and\ \bibinfo {author} {\bibfnamefont {D.}~\bibnamefont {Genov}},\ }\bibfield  {title} {\bibinfo {title} {Trapping light by mimicking gravitational lensing},\ }in\ \href@noop {} {\emph {\bibinfo {booktitle} {Frontiers in Optics}}}\ (\bibinfo {organization} {Optica Publishing Group},\ \bibinfo {year} {2014})\ pp.\ \bibinfo {pages} {FTu5D--1}\BibitemShut {NoStop}%
\bibitem [{\citenamefont {Koke}\ \emph {et~al.}(2016)\citenamefont {Koke}, \citenamefont {Noh},\ and\ \citenamefont {Angelakis}}]{koke2016dirac}%
  \BibitemOpen
  \bibfield  {author} {\bibinfo {author} {\bibfnamefont {C.}~\bibnamefont {Koke}}, \bibinfo {author} {\bibfnamefont {C.}~\bibnamefont {Noh}},\ and\ \bibinfo {author} {\bibfnamefont {D.~G.}\ \bibnamefont {Angelakis}},\ }\bibfield  {title} {\bibinfo {title} {Dirac equation in 2-dimensional curved spacetime, particle creation, and coupled waveguide arrays},\ }\href@noop {} {\bibfield  {journal} {\bibinfo  {journal} {Annals of Physics}\ }\textbf {\bibinfo {volume} {374}},\ \bibinfo {pages} {162} (\bibinfo {year} {2016})}\BibitemShut {NoStop}%
\bibitem [{\citenamefont {Pedernales}\ \emph {et~al.}(2018)\citenamefont {Pedernales}, \citenamefont {Beau}, \citenamefont {Pittman}, \citenamefont {Egusquiza}, \citenamefont {Lamata}, \citenamefont {Solano},\ and\ \citenamefont {del Campo}}]{pedernales2018dirac}%
  \BibitemOpen
  \bibfield  {author} {\bibinfo {author} {\bibfnamefont {J.~S.}\ \bibnamefont {Pedernales}}, \bibinfo {author} {\bibfnamefont {M.}~\bibnamefont {Beau}}, \bibinfo {author} {\bibfnamefont {S.~M.}\ \bibnamefont {Pittman}}, \bibinfo {author} {\bibfnamefont {I.~L.}\ \bibnamefont {Egusquiza}}, \bibinfo {author} {\bibfnamefont {L.}~\bibnamefont {Lamata}}, \bibinfo {author} {\bibfnamefont {E.}~\bibnamefont {Solano}},\ and\ \bibinfo {author} {\bibfnamefont {A.}~\bibnamefont {del Campo}},\ }\bibfield  {title} {\bibinfo {title} {Dirac equation in (1+ 1)-dimensional curved spacetime and the multiphoton quantum rabi model},\ }\href@noop {} {\bibfield  {journal} {\bibinfo  {journal} {Physical Review Letters}\ }\textbf {\bibinfo {volume} {120}},\ \bibinfo {pages} {160403} (\bibinfo {year} {2018})}\BibitemShut {NoStop}%
\bibitem [{\citenamefont {Yan}\ \emph {et~al.}(2018)\citenamefont {Yan}, \citenamefont {Krantz}, \citenamefont {Sung}, \citenamefont {Kjaergaard}, \citenamefont {Campbell}, \citenamefont {Orlando}, \citenamefont {Gustavsson},\ and\ \citenamefont {Oliver}}]{yan2018tunable}%
  \BibitemOpen
  \bibfield  {author} {\bibinfo {author} {\bibfnamefont {F.}~\bibnamefont {Yan}}, \bibinfo {author} {\bibfnamefont {P.}~\bibnamefont {Krantz}}, \bibinfo {author} {\bibfnamefont {Y.}~\bibnamefont {Sung}}, \bibinfo {author} {\bibfnamefont {M.}~\bibnamefont {Kjaergaard}}, \bibinfo {author} {\bibfnamefont {D.~L.}\ \bibnamefont {Campbell}}, \bibinfo {author} {\bibfnamefont {T.~P.}\ \bibnamefont {Orlando}}, \bibinfo {author} {\bibfnamefont {S.}~\bibnamefont {Gustavsson}},\ and\ \bibinfo {author} {\bibfnamefont {W.~D.}\ \bibnamefont {Oliver}},\ }\bibfield  {title} {\bibinfo {title} {Tunable coupling scheme for implementing high-fidelity two-qubit gates},\ }\href@noop {} {\bibfield  {journal} {\bibinfo  {journal} {Physical Review Applied}\ }\textbf {\bibinfo {volume} {10}},\ \bibinfo {pages} {054062} (\bibinfo {year} {2018})}\BibitemShut {NoStop}%
\bibitem [{\citenamefont {Sung}\ \emph {et~al.}(2021)\citenamefont {Sung}, \citenamefont {Ding}, \citenamefont {Braum{\"u}ller}, \citenamefont {Veps{\"a}l{\"a}inen}, \citenamefont {Kannan}, \citenamefont {Kjaergaard}, \citenamefont {Greene}, \citenamefont {Samach}, \citenamefont {McNally}, \citenamefont {Kim} \emph {et~al.}}]{sung2021realization}%
  \BibitemOpen
  \bibfield  {author} {\bibinfo {author} {\bibfnamefont {Y.}~\bibnamefont {Sung}}, \bibinfo {author} {\bibfnamefont {L.}~\bibnamefont {Ding}}, \bibinfo {author} {\bibfnamefont {J.}~\bibnamefont {Braum{\"u}ller}}, \bibinfo {author} {\bibfnamefont {A.}~\bibnamefont {Veps{\"a}l{\"a}inen}}, \bibinfo {author} {\bibfnamefont {B.}~\bibnamefont {Kannan}}, \bibinfo {author} {\bibfnamefont {M.}~\bibnamefont {Kjaergaard}}, \bibinfo {author} {\bibfnamefont {A.}~\bibnamefont {Greene}}, \bibinfo {author} {\bibfnamefont {G.~O.}\ \bibnamefont {Samach}}, \bibinfo {author} {\bibfnamefont {C.}~\bibnamefont {McNally}}, \bibinfo {author} {\bibfnamefont {D.}~\bibnamefont {Kim}}, \emph {et~al.},\ }\bibfield  {title} {\bibinfo {title} {Realization of high-fidelity cz and zz-free iswap gates with a tunable coupler},\ }\href@noop {} {\bibfield  {journal} {\bibinfo  {journal} {Physical Review X}\ }\textbf {\bibinfo {volume} {11}},\ \bibinfo {pages} {021058} (\bibinfo {year} {2021})}\BibitemShut {NoStop}%
\bibitem [{\citenamefont {Tian}\ and\ \citenamefont {Du}(2019)}]{tian2019analogue}%
  \BibitemOpen
  \bibfield  {author} {\bibinfo {author} {\bibfnamefont {Z.}~\bibnamefont {Tian}}\ and\ \bibinfo {author} {\bibfnamefont {J.}~\bibnamefont {Du}},\ }\bibfield  {title} {\bibinfo {title} {Analogue hawking radiation and quantum soliton evaporation in a superconducting circuit},\ }\href@noop {} {\bibfield  {journal} {\bibinfo  {journal} {The European Physical Journal C}\ }\textbf {\bibinfo {volume} {79}},\ \bibinfo {pages} {994} (\bibinfo {year} {2019})}\BibitemShut {NoStop}%
\bibitem [{\citenamefont {Shi}\ \emph {et~al.}(2023)\citenamefont {Shi}, \citenamefont {Yang}, \citenamefont {Xiang}, \citenamefont {Ge}, \citenamefont {Li}, \citenamefont {Wang}, \citenamefont {Huang}, \citenamefont {Tian}, \citenamefont {Song}, \citenamefont {Zheng} \emph {et~al.}}]{shi2023quantum}%
  \BibitemOpen
  \bibfield  {author} {\bibinfo {author} {\bibfnamefont {Y.-H.}\ \bibnamefont {Shi}}, \bibinfo {author} {\bibfnamefont {R.-Q.}\ \bibnamefont {Yang}}, \bibinfo {author} {\bibfnamefont {Z.}~\bibnamefont {Xiang}}, \bibinfo {author} {\bibfnamefont {Z.-Y.}\ \bibnamefont {Ge}}, \bibinfo {author} {\bibfnamefont {H.}~\bibnamefont {Li}}, \bibinfo {author} {\bibfnamefont {Y.-Y.}\ \bibnamefont {Wang}}, \bibinfo {author} {\bibfnamefont {K.}~\bibnamefont {Huang}}, \bibinfo {author} {\bibfnamefont {Y.}~\bibnamefont {Tian}}, \bibinfo {author} {\bibfnamefont {X.}~\bibnamefont {Song}}, \bibinfo {author} {\bibfnamefont {D.}~\bibnamefont {Zheng}}, \emph {et~al.},\ }\bibfield  {title} {\bibinfo {title} {Quantum simulation of hawking radiation and curved spacetime with a superconducting on-chip black hole},\ }\href@noop {} {\bibfield  {journal} {\bibinfo  {journal} {Nature Communications}\ }\textbf {\bibinfo {volume} {14}},\ \bibinfo {pages} {3263} (\bibinfo {year} {2023})}\BibitemShut {NoStop}%
\bibitem [{\citenamefont {Sabín}(2016)}]{Carlos2016Mapping}%
  \BibitemOpen
  \bibfield  {author} {\bibinfo {author} {\bibfnamefont {C.}~\bibnamefont {Sabín}},\ }\bibfield  {title} {\bibinfo {title} {Mapping curved spacetimes into dirac spinors},\ }\href@noop {} {\bibfield  {journal} {\bibinfo  {journal} {Scientific Reports}\ }\textbf {\bibinfo {volume} {7}} (\bibinfo {year} {2016})}\BibitemShut {NoStop}%
\bibitem [{\citenamefont {Sab{\'i}n}\ \emph {et~al.}(2012)\citenamefont {Sab{\'i}n}, \citenamefont {Casanova}, \citenamefont {Garc{\'i}a-Ripoll}, \citenamefont {Lamata}, \citenamefont {Solano},\ and\ \citenamefont {Le{\'o}n}}]{Sabn2012EncodingRP}%
  \BibitemOpen
  \bibfield  {author} {\bibinfo {author} {\bibfnamefont {C.}~\bibnamefont {Sab{\'i}n}}, \bibinfo {author} {\bibfnamefont {J.}~\bibnamefont {Casanova}}, \bibinfo {author} {\bibfnamefont {J.~J.}\ \bibnamefont {Garc{\'i}a-Ripoll}}, \bibinfo {author} {\bibfnamefont {L.}~\bibnamefont {Lamata}}, \bibinfo {author} {\bibfnamefont {E.}~\bibnamefont {Solano}},\ and\ \bibinfo {author} {\bibfnamefont {J.}~\bibnamefont {Le{\'o}n}},\ }\bibfield  {title} {\bibinfo {title} {Encoding relativistic potential dynamics into free evolution},\ }\href@noop {} {\bibfield  {journal} {\bibinfo  {journal} {Physical Review A}\ }\textbf {\bibinfo {volume} {85}},\ \bibinfo {pages} {052301} (\bibinfo {year} {2012})}\BibitemShut {NoStop}%
\bibitem [{\citenamefont {Yang}\ \emph {et~al.}(2020)\citenamefont {Yang}, \citenamefont {Liu}, \citenamefont {Zhu}, \citenamefont {Luo},\ and\ \citenamefont {Cai}}]{yang2020simulating}%
  \BibitemOpen
  \bibfield  {author} {\bibinfo {author} {\bibfnamefont {R.-Q.}\ \bibnamefont {Yang}}, \bibinfo {author} {\bibfnamefont {H.}~\bibnamefont {Liu}}, \bibinfo {author} {\bibfnamefont {S.}~\bibnamefont {Zhu}}, \bibinfo {author} {\bibfnamefont {L.}~\bibnamefont {Luo}},\ and\ \bibinfo {author} {\bibfnamefont {R.-G.}\ \bibnamefont {Cai}},\ }\bibfield  {title} {\bibinfo {title} {Simulating quantum field theory in curved spacetime with quantum many-body systems},\ }\href@noop {} {\bibfield  {journal} {\bibinfo  {journal} {Physical Review Research}\ }\textbf {\bibinfo {volume} {2}},\ \bibinfo {pages} {023107} (\bibinfo {year} {2020})}\BibitemShut {NoStop}%
\bibitem [{\citenamefont {Deger}\ \emph {et~al.}(2023)\citenamefont {Deger}, \citenamefont {Horner},\ and\ \citenamefont {Pachos}}]{deger2023ads}%
  \BibitemOpen
  \bibfield  {author} {\bibinfo {author} {\bibfnamefont {A.}~\bibnamefont {Deger}}, \bibinfo {author} {\bibfnamefont {M.~D.}\ \bibnamefont {Horner}},\ and\ \bibinfo {author} {\bibfnamefont {J.~K.}\ \bibnamefont {Pachos}},\ }\bibfield  {title} {\bibinfo {title} {Ads/cft correspondence with a three-dimensional black hole simulator},\ }\href@noop {} {\bibfield  {journal} {\bibinfo  {journal} {Physical Review B}\ }\textbf {\bibinfo {volume} {108}},\ \bibinfo {pages} {155124} (\bibinfo {year} {2023})}\BibitemShut {NoStop}%
\bibitem [{\citenamefont {Sinha}\ and\ \citenamefont {Roychoudhury}(1994)}]{1994Dirac}%
  \BibitemOpen
  \bibfield  {author} {\bibinfo {author} {\bibfnamefont {A.}~\bibnamefont {Sinha}}\ and\ \bibinfo {author} {\bibfnamefont {R.}~\bibnamefont {Roychoudhury}},\ }\bibfield  {title} {\bibinfo {title} {Dirac equation in (1+1)-dimensional curved space-time},\ }\href@noop {} {\bibfield  {journal} {\bibinfo  {journal} {International Journal of Theoretical Physics}\ }\textbf {\bibinfo {volume} {33}},\ \bibinfo {pages} {1511} (\bibinfo {year} {1994})}\BibitemShut {NoStop}%
\bibitem [{\citenamefont {Mann}\ \emph {et~al.}(1991)\citenamefont {Mann}, \citenamefont {Morsink}, \citenamefont {Sikkema},\ and\ \citenamefont {Steele}}]{mann1991semiclassical}%
  \BibitemOpen
  \bibfield  {author} {\bibinfo {author} {\bibfnamefont {R.}~\bibnamefont {Mann}}, \bibinfo {author} {\bibfnamefont {S.}~\bibnamefont {Morsink}}, \bibinfo {author} {\bibfnamefont {A.}~\bibnamefont {Sikkema}},\ and\ \bibinfo {author} {\bibfnamefont {T.}~\bibnamefont {Steele}},\ }\bibfield  {title} {\bibinfo {title} {Semiclassical gravity in 1+ 1 dimensions},\ }\href@noop {} {\bibfield  {journal} {\bibinfo  {journal} {Physical Review D}\ }\textbf {\bibinfo {volume} {43}},\ \bibinfo {pages} {3948} (\bibinfo {year} {1991})}\BibitemShut {NoStop}%
\bibitem [{\citenamefont {Morsink}\ and\ \citenamefont {Mann}(1991)}]{morsink1991black}%
  \BibitemOpen
  \bibfield  {author} {\bibinfo {author} {\bibfnamefont {S.~M.}\ \bibnamefont {Morsink}}\ and\ \bibinfo {author} {\bibfnamefont {R.~B.}\ \bibnamefont {Mann}},\ }\bibfield  {title} {\bibinfo {title} {Black hole radiation of dirac particles in 1+ 1 dimensions},\ }\href@noop {} {\bibfield  {journal} {\bibinfo  {journal} {Classical and Quantum Gravity}\ }\textbf {\bibinfo {volume} {8}},\ \bibinfo {pages} {2257} (\bibinfo {year} {1991})}\BibitemShut {NoStop}%
\bibitem [{\citenamefont {Barouch}\ \emph {et~al.}(1970)\citenamefont {Barouch}, \citenamefont {Mccoy},\ and\ \citenamefont {Dresden}}]{1970Statistical}%
  \BibitemOpen
  \bibfield  {author} {\bibinfo {author} {\bibfnamefont {E.}~\bibnamefont {Barouch}}, \bibinfo {author} {\bibfnamefont {B.~M.}\ \bibnamefont {Mccoy}},\ and\ \bibinfo {author} {\bibfnamefont {M.}~\bibnamefont {Dresden}},\ }\bibfield  {title} {\bibinfo {title} {Statistical mechanics of the model. i},\ }\href@noop {} {\bibfield  {journal} {\bibinfo  {journal} {Physical Review A}\ }\textbf {\bibinfo {volume} {2}},\ \bibinfo {pages} {1075} (\bibinfo {year} {1970})}\BibitemShut {NoStop}%
\bibitem [{\citenamefont {Barouch}\ and\ \citenamefont {McCoy}(1971)}]{barouch1971statistical}%
  \BibitemOpen
  \bibfield  {author} {\bibinfo {author} {\bibfnamefont {E.}~\bibnamefont {Barouch}}\ and\ \bibinfo {author} {\bibfnamefont {B.~M.}\ \bibnamefont {McCoy}},\ }\bibfield  {title} {\bibinfo {title} {Statistical mechanics of the x y model. ii. spin-correlation functions},\ }\href@noop {} {\bibfield  {journal} {\bibinfo  {journal} {Physical Review A}\ }\textbf {\bibinfo {volume} {3}},\ \bibinfo {pages} {786} (\bibinfo {year} {1971})}\BibitemShut {NoStop}%
\bibitem [{\citenamefont {Damour}\ and\ \citenamefont {Ruffini}(1976)}]{damour1976black}%
  \BibitemOpen
  \bibfield  {author} {\bibinfo {author} {\bibfnamefont {T.}~\bibnamefont {Damour}}\ and\ \bibinfo {author} {\bibfnamefont {R.}~\bibnamefont {Ruffini}},\ }\bibfield  {title} {\bibinfo {title} {Black-hole evaporation in the klein-sauter-heisenberg-euler formalism},\ }\href@noop {} {\bibfield  {journal} {\bibinfo  {journal} {Physical Review D}\ }\textbf {\bibinfo {volume} {14}},\ \bibinfo {pages} {332} (\bibinfo {year} {1976})}\BibitemShut {NoStop}%
\bibitem [{\citenamefont {Simpson}\ and\ \citenamefont {Visser}(2019)}]{simpson2019black}%
  \BibitemOpen
  \bibfield  {author} {\bibinfo {author} {\bibfnamefont {A.}~\bibnamefont {Simpson}}\ and\ \bibinfo {author} {\bibfnamefont {M.}~\bibnamefont {Visser}},\ }\bibfield  {title} {\bibinfo {title} {Black-bounce to traversable wormhole},\ }\href@noop {} {\bibfield  {journal} {\bibinfo  {journal} {Journal of Cosmology and Astroparticle Physics}\ }\textbf {\bibinfo {volume} {2019}}\bibinfo  {number} { (02)},\ \bibinfo {pages} {042}}\BibitemShut {NoStop}%
\bibitem [{\citenamefont {Painlev{\'e}}(1921)}]{painleve1921comptes}%
  \BibitemOpen
\bibfield  {number} {  }\bibfield  {author} {\bibinfo {author} {\bibfnamefont {P.}~\bibnamefont {Painlev{\'e}}},\ }\bibfield  {title} {\bibinfo {title} {Comptes rendus de l’academie des sciences},\ }\href@noop {} {\bibfield  {journal} {\bibinfo  {journal} {Serie I (Mathematique)}\ }\textbf {\bibinfo {volume} {173}},\ \bibinfo {pages} {677} (\bibinfo {year} {1921})}\BibitemShut {NoStop}%
\bibitem [{\citenamefont {et~al}(2008)}]{liuliao2008Nature}%
  \BibitemOpen
  \bibfield  {author} {\bibinfo {author} {\bibfnamefont {L.~L.}\ \bibnamefont {et~al}},\ }\href@noop {} {\emph {\bibinfo {title} {The Nature of Black Hole and Time(in chinese)}}}\ (\bibinfo  {publisher} {Peking University Press},\ \bibinfo {year} {2008})\BibitemShut {NoStop}%
\bibitem [{\citenamefont {Sch{\"u}tzhold}\ and\ \citenamefont {Unruh}(2002)}]{schutzhold2002gravity}%
  \BibitemOpen
  \bibfield  {author} {\bibinfo {author} {\bibfnamefont {R.}~\bibnamefont {Sch{\"u}tzhold}}\ and\ \bibinfo {author} {\bibfnamefont {W.~G.}\ \bibnamefont {Unruh}},\ }\bibfield  {title} {\bibinfo {title} {Gravity wave analogues of black holes},\ }\href@noop {} {\bibfield  {journal} {\bibinfo  {journal} {Physical Review D}\ }\textbf {\bibinfo {volume} {66}},\ \bibinfo {pages} {044019} (\bibinfo {year} {2002})}\BibitemShut {NoStop}%
\bibitem [{\citenamefont {Weinfurtner}\ \emph {et~al.}(2013)\citenamefont {Weinfurtner}, \citenamefont {Tedford}, \citenamefont {Penrice}, \citenamefont {Unruh},\ and\ \citenamefont {Lawrence}}]{weinfurtner2013classical}%
  \BibitemOpen
  \bibfield  {author} {\bibinfo {author} {\bibfnamefont {S.}~\bibnamefont {Weinfurtner}}, \bibinfo {author} {\bibfnamefont {E.~W.}\ \bibnamefont {Tedford}}, \bibinfo {author} {\bibfnamefont {M.~C.}\ \bibnamefont {Penrice}}, \bibinfo {author} {\bibfnamefont {W.~G.}\ \bibnamefont {Unruh}},\ and\ \bibinfo {author} {\bibfnamefont {G.~A.}\ \bibnamefont {Lawrence}},\ }\bibfield  {title} {\bibinfo {title} {Classical aspects of hawking radiation verified in analogue gravity experiment},\ }\href@noop {} {\bibfield  {journal} {\bibinfo  {journal} {Analogue gravity phenomenology: Analogue spacetimes and horizons, from theory to experiment}\ ,\ \bibinfo {pages} {167}} (\bibinfo {year} {2013})}\BibitemShut {NoStop}%
\bibitem [{\citenamefont {Morris}\ \emph {et~al.}(1988)\citenamefont {Morris}, \citenamefont {Thorne},\ and\ \citenamefont {Yurtsever}}]{1988Wormholes}%
  \BibitemOpen
  \bibfield  {author} {\bibinfo {author} {\bibfnamefont {M.~S.}\ \bibnamefont {Morris}}, \bibinfo {author} {\bibfnamefont {K.~S.}\ \bibnamefont {Thorne}},\ and\ \bibinfo {author} {\bibfnamefont {U.}~\bibnamefont {Yurtsever}},\ }\bibfield  {title} {\bibinfo {title} {Wormholes, time machines, and the weak energy condition},\ }\href@noop {} {\bibfield  {journal} {\bibinfo  {journal} {Physical Review Letters}\ }\textbf {\bibinfo {volume} {61}},\ \bibinfo {pages} {1446–1449} (\bibinfo {year} {1988})}\BibitemShut {NoStop}%
\bibitem [{\citenamefont {Morris}\ and\ \citenamefont {Thorne}(1988)}]{1988WormholesII}%
  \BibitemOpen
  \bibfield  {author} {\bibinfo {author} {\bibfnamefont {M.~S.}\ \bibnamefont {Morris}}\ and\ \bibinfo {author} {\bibfnamefont {K.~S.}\ \bibnamefont {Thorne}},\ }\bibfield  {title} {\bibinfo {title} {Wormholes in spacetime and their use for interstellar travel: A tool to teaching general relativity},\ }\href@noop {} {\bibfield  {journal} {\bibinfo  {journal} {American Journal of Physics}\ }\textbf {\bibinfo {volume} {56}} (\bibinfo {year} {1988})}\BibitemShut {NoStop}%
\bibitem [{\citenamefont {Lobo}\ \emph {et~al.}(2020)\citenamefont {Lobo}, \citenamefont {Simpson},\ and\ \citenamefont {Visser}}]{lobo2020dynamic}%
  \BibitemOpen
  \bibfield  {author} {\bibinfo {author} {\bibfnamefont {F.~S.}\ \bibnamefont {Lobo}}, \bibinfo {author} {\bibfnamefont {A.}~\bibnamefont {Simpson}},\ and\ \bibinfo {author} {\bibfnamefont {M.}~\bibnamefont {Visser}},\ }\bibfield  {title} {\bibinfo {title} {Dynamic thin-shell black-bounce traversable wormholes},\ }\href@noop {} {\bibfield  {journal} {\bibinfo  {journal} {Physical Review D}\ }\textbf {\bibinfo {volume} {101}},\ \bibinfo {pages} {124035} (\bibinfo {year} {2020})}\BibitemShut {NoStop}%
\end{thebibliography}
%

\end{document}